\documentclass{statsoc}

\usepackage{float}                      
\usepackage{natbib}                     
\usepackage{graphicx}					
\graphicspath{ {figures/}{figures/eps/}{figures/pdf/} }
\usepackage{fancyhdr}                   
\usepackage{url}                        
\usepackage[inactive]{srcltx}		 	
\usepackage{relsize}                    
\usepackage{booktabs}                   
\usepackage{mathrsfs}                   

\usepackage[a4paper,margin=1.3in]{geometry}
\usepackage[textwidth=8em,textsize=small]{todonotes}
\usepackage{amsmath,amssymb,setspace}
\usepackage{longtable}

\newif\ifdraft
\drafttrue
\ifdraft
\newcommand{\xtai}[1]{\textcolor{blue}{[[Xiao Hui says: #1]]}}
\else
\newcommand{\xtai}[1]{}
\fi

\usepackage{subcaption}
\usepackage{microtype} 
\usepackage{algorithm} 
\usepackage{algpseudocode} 
\usepackage{enumitem}
\usepackage[stable]{footmisc}
\usepackage{tikz}

\title[Automatically matching topographical measurements of cartridge cases]{Automatically matching topographical measurements of cartridge cases using a record linkage framework}
\author[Xiao Hui Tai {\it et al.}]{Xiao Hui Tai}
\address{U.C. Berkeley,
	Berkeley,
	USA.}
\email{xtai@berkeley.edu}
\author{William F. Eddy}
\address{Carnegie Mellon University,
	Pittsburgh,
	USA.}

\begin{document}

\begin{abstract}
Firing a gun leaves marks on cartridge cases which purportedly uniquely identify the gun. Firearms examiners typically use a visual examination to evaluate if two cartridge cases were fired from the same gun, and this is a subjective process that has come under scrutiny. Matching can be done in a more reproducible manner using automated algorithms. 
In this paper, we develop methodology to compare topographical measurements of cartridge cases. We demonstrate the use of a record linkage framework in this context. 
We compare performance using topographical measurements to older reflectance microscopy images, investigating the extent to which the former produce more accurate comparisons. Using a diverse collection of images of over 1,100 cartridge cases, 
we find that overall performance is generally improved using topographical data. 
Some subsets of the data achieve almost perfect predictive performance in terms of precision and recall, while some produce extremely poor performance. 
Further work needs to be done to assess if examiners face similar difficulties on certain gun and ammunition combinations. 
For automatic methods, a fuller investigation into their fairness and robustness is necessary before they can be deployed in practice. 

	
	


\end{abstract}


\section{Introduction}
\label{sec:intro}
When a crime is committed, the perpetrator almost invariably leaves behind traces of evidence, which could take various forms, for example DNA, fingerprints, bullets, cartridge cases, shoeprints or digital evidence. Forensic matching involves comparing pairs of samples, to infer if they originated from the same source. The underlying assumption is that pieces of evidence have identifiable characteristics that can be traced back to their source.

\begin{figure}[!ht]
	\caption{\label{fig:mechanism}
		Mechanism by which marks are left on cartridge cases.}
	
	\centering
	\begin{subfigure}[t]{0.22\textwidth}
		\centering
		\includegraphics[width=\columnwidth]{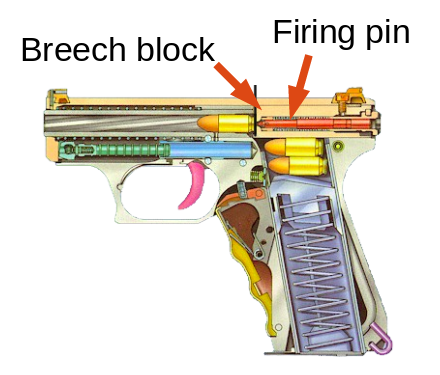}
		\caption{Gun that is about to be fired, showing the internal parts. 
		\label{fig:gun}}
	\end{subfigure}
	~
	\begin{subfigure}[t]{0.74\textwidth}
		\centering
		\includegraphics[width = \columnwidth]{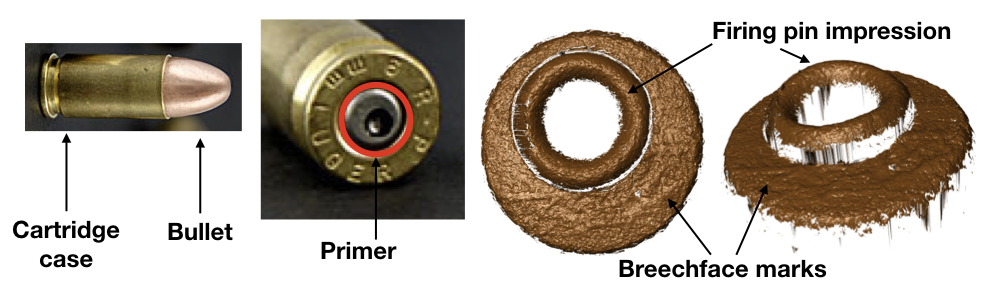}
		\caption{On the far left is a cartridge before firing. In the middle left is the bottom surface of a cartridge case after firing. On the right are images of such a bottom surface, taken using a confocal microscope. The rightmost image shows the image rotated on a horizontal axis. 
	\label{fig:marks}}
	\end{subfigure}	
	
\end{figure}

In the case of firearms, firing a gun leaves marks on the bottom surface of the cartridge case. The mechanism by which this happens is illustrated in Figure \ref{fig:mechanism}. Notice in Figure \ref{fig:mechanism}\subref{fig:gun} that the bottom surface of the cartridge is in contact with the breech block of the gun and the firing pin. During the firing process the cartridge is hit by the firing pin, which causes it to break up into two components: the bullet which goes out the barrel, and the cartridge case that is ejected from the side. This process leaves at least two kinds of marks, seen in Figure \ref{fig:mechanism}\subref{fig:marks}. The firing pin hits the cartridge, leaving a firing pin impression, and the breech block presses against the primer of the cartridge, leaving breechface marks. The breech block is harder than the primer, and may have microscopic defects or patterns that are impressed on the primer. This paper focuses exclusively on breechface marks. 


Law enforcement officers routinely collect guns and cartridge cases from crime scenes, because of their potential usefulness in investigations. The Bureau of Alcohol, Tobacco, Firearms and Explosives (ATF) in the United States maintains a national database called the National Integrated Ballistics Information Network (NIBIN). This contains around 3.3 million images of cartridge cases retrieved from crime scenes, and an additional 12.7 million test fires (as of May 2018) \citep{Bureau-of-Alcohol:2018aa}. In current practice, cartridge cases may be entered into NIBIN through a computer-based platform called the Integrated Ballistics Identification System (IBIS), which was developed and is maintained by Ultra Electronics Forensic Technology Inc. This platform captures an image of the retrieved cartridge case and runs a proprietary search against some user-selected subset of the database (for example images from a certain geographical location). This returns a list of top ranked potential matches. Firearms examiners may then select a subset that they find to be promising. They may then locate the physical cartridge cases associated with these images, to be examined under a comparison microscope. The final determination of a match or non-match is made by the examiner, based on whether there is ``sufficient agreement'' between the marks \citep{AFTE1992}. The examiner may then bring this evidence to court. It is the examiner that attributes any evidence to a particular source; the NIBIN system is simply used as an investigative tool to generate leads. It is up to the discretion of firearms examiners whether or not to enter retrieved cartridge cases into NIBIN. In addition, the NIBIN database is proprietary and unavailable for study.

The current system has been fraught with controversy. Historically, forensic methods in general have been vetted by the legal system, as opposed to the scientific community. Beginning in the 1990s, exonerations due to DNA evidence revealed problems in many forensic science disciplines \citep{Bell2018}. Examiners have been found to have overstated forensic results, leading (at least in part) to wrongful convictions \citep{Murphy:2019aa}. A 2009 National Academy of Sciences report \citep{Council2009} called for an overhaul of the forensic science system. With respect to forensic matching, it stated that with the exception of nuclear DNA analysis, disciplines had neither scientific support nor a proper quantification of error rates or limitations. In 2016, the President's Council of Advisors on Science and Technology (PCAST) followed up with an independent report \citep{PCAST2016}, specifically addressing any scientific developments in the various pattern matching disciplines. For firearms analysis, there were concerns that there had been insufficient studies establishing the reliability of conclusions made by examiners, and the associated error rates had not been adequately determined. The report suggested two directions for the future. The first is to ``continue to improve firearms analysis as a subjective method,'' by conducting further studies, and the second is to ``convert firearms analysis from a subjective method to an objective method.''

Commercial and academic groups have made some progress with respect to the second recommendation. For breechface marks, examples of related work include \cite{Song2018}, \cite{Roth2015}, \cite{Riva2014}, \cite{Thumwarin2008} and \cite{Geradts2001}. The general methodology is as follows. First select the relevant breechface marks, using some form of manual input. Then use various image processing steps to extract relevant features, or derive a similarity score directly. For example, \cite{Song2018} split the breechface area into a grid of cells and aligns them independently, producing a similarity score that is defined as the number of matching cells. A pre-specified cutoff is used for classifying a pair as matched, in other words coming from the same source. \cite{Roth2015} use a supervised classification method (gradient boosting) to classify pairs of images as being matched or not. \cite{Giudice:2019aa} employ a slightly different strategy, developing a fully automatic method that does not require any pre-processing.

On the commercial side, Ultra Electronics Forensic Technology Inc. maintains the system used in NIBIN. The software extracts a numerical signature from each region of interest and does a database search using these signatures \citep{UEFT2018}. The methodology is proprietary and few details are known. Cadre Forensics is another manufacturer of microscopes. They extract geometric feature points that ``a trained firearms examiner would identify,'' such as ridges, peaks, gouges, and concavities. These features are then aligned, as a set, to the features from another image. Logistic regression is used to produce a similarity score, with covariates such as number of matched features as input \citep{Lilien2017}.

The shortcomings of currently available methods are that none of them are both fully automatic and open source. Automatic methods ensure that results are objective and reproducible, and open source software allows proper testing and validation. In this paper, we contribute methodology for comparing topographical measurements of cartridge cases, that is both open source and fully automatic, allowing for a transparent evaluation. This builds on work in \cite{Tai2018}, which focuses on reflectance images as opposed to topographical measurements (more details in Section \ref{sec:data}), and \cite{Tai:2018aa}, which proposes a framework for developing forensic matching methods. Software is available at \url{https://github.com/xhtai/cartridges3D}. We apply a record linkage framework to forensic matching, demonstrating the principled use of statistical methods in the forensic sciences, at the same time contributing an application of record linkage methods that is non-traditional. We conduct a comprehensive analysis involving over 1,100 images, enabling us to compare performance across different firearms and ammunition. Further, we do a detailed comparison of results produced using older reflectance microscopy technology versus topographical measurements. All these (and more; see Section \ref{sec:discussion}) are essential steps before such automated methods can be deployed in actual criminal cases, or determined to be an improvement over the current system. At the same time, more work needs to be done to compare the results of automatic methods to examiner conclusions. If the two are related, results based on automatic methods might shed some light on examiner error rates over a large number of examples, and provide insight into comparisons that examiners may struggle with. 

The rest of the paper is organized as follows. In Section \ref{sec:data} we first introduce the data and types of digital representations that are possible with cartridge cases. Section \ref{sec:method} presents the methodology in the context of a record linkage problem. Results are in Section \ref{sec:results}, and Section \ref{sec:discussion} concludes.

\section{Data}
\label{sec:data}

Current technology allows for digital representations of fired cartridge cases to be captured and stored for later comparison. Reflectance microscopes measure the amount of light reflected off the object's surface, producing grayscale images. In the literature these are often called 2D images. Newer technology in the form of 3D profilers (such as confocal microscopes) measure surface contours directly, producing topographical measurements. These are often termed 3D images, although both these types of data come in the form of 2D matrices, with entries being either reflectance or depth values. For simplicity, in the remainder of this paper we refer to the two types of representations as 2D and 3D images.

In recent years the National Institute of Standards and Technology (NIST) has advocated the use of 3D images because of their insensitivity to lighting conditions and traceability to the International System of Units (SI) unit of length \citep{Song2012}. The SI system comprises units of measurement built on base units that have precise standards of measurements. Specifically, base units are derived from invariant constants of nature (that can be observed and measured with great accuracy), and one physical artifact. This means that measurements of cartridge cases using any instrument can be compared to a known standard, and instruments can be calibrated to this known standard to assess and ensure precision. 

The focus of this paper is on developing methods for 3D data. We also compare the results from comparisons made using 2D and 3D data, to investigate if purportedly more accurate measurements translate into better comparison results. In the literature, methods are typically developed for either 2D or 3D data, with limited direct comparisons made between the two. 


We use data from NIST's Ballistics Toolmark Research Database (\url{https://tsapps.nist.gov/NRBTD}). This is an open-access research database of bullet and cartridge case toolmark data, which to our knowledge, is the largest publicly available collection of reference data. Although a national database of firearms data exists (NIBIN), these data are not publicly available. The NIST database contains images originating from studies conducted by various groups in the firearm and toolmark community. 
Almost all of the currently available data are of cartridge cases that were sent to NIST for imaging, but the website also allows users to upload their own data in a standardized format. We study a subset of 1,158 of these images. We limit our analysis to measurements made by NIST, using disc scanning confocal microscopy. The technology for measuring surface topographies is relatively new, and as far as we know, there are no standardized procedures on how these images should be captured. For this paper, we begin with NIST-collected measurements made using a single method of measurement, for the sake of uniformity and ease of comparison. Future work could involve examining the variability due to instrument, method of measurement, and so forth.

The database contains both 2D and 3D images. The various data sets used are summarized in Table \ref{tab:firearmsData}, with each data set containing images from a single study. 
 Among these data are sets involving consecutively manufactured pistol slides, a large number of firings (termed persistence studies because they investigate the persistence of marks), as well as a combination of different makes and models of guns and ammunition. Gun manufacturers include Glock, Hi-Point, Ruger, Sig Sauer, and Smith \& Wesson, and ammunition brands include CCI, Federal, PMC, Remington, Speer, Wolf and Winchester. Metadata available for download provide additional information such as study details, the type of firing pin, material of the primer, etc.


\fussy

\begin{center}
	\captionof{table}{Summary of data used from NIST's Ballistics Toolmark Research Database. 2D and 3D data are available for all studies listed, with the exception of CTS, where only 3D images exist. Note that for Todd Weller 95 cartridge cases were imaged in 3D but only 50 were imaged in 2D.}
 \label{tab:firearmsData} 
	\centering
	\hspace*{-1.7cm}\begin{tabular}{lclcclc} 
		\hline 
		Study & Cartridge & Firearm & Number & Slides per  & Cartridge & Test fires per  \\
		& cases & & of firearms & firearm &  & firearm/slide \\
		\hline \hline
		Cary Wong & 91 & Ruger P89 & 1 & 1 & Winchester & 91 \\
		\hline
		CTS & 74 & Ruger P94DC & 1 & 1 & Federal & 44 \\
		&  & Ruger P91DC & 1 & 1 & Federal  & 18 \\
		&  & S\&W SW40VE & 1 & 1 & Federal  & 12 \\
		\hline
		De Kinder       &  70    &  Sig Sauer P226 & 10  &     1 &      Remington &  2 \\
		& & & & & CCI & 1\\
		& & & & & Wolf & 1\\
		& & & & & Winchester & 1\\
		& & & & & Speer & 1\\
		& & & & & Federal & 1\\
		\hline
		Thomas Fadul   &  40  &   Ruger P95PR15 &  1 &       10 &     Federal   &  3-5 \\
		\hline
		Hamby & 30 & Hi-Point C9 & 1 & 10 & Remington & 3 \\
		\hline
		Kong & 36 & S\&W 10-10 & 12 & 1 & Fiocchi & 3 \\
		\hline
		Laura Lightstone & 30 & S\&W 40VE & 1 & 10 & PMC & 3 \\
		\hline		
		NIST Ballistics   & 144 & Ruger P95D  & 4      &  1    &   Remington &  3 \\
		Imaging  & & & & &  Winchester & 3 \\
		Database & & & & &  Speer & 3 \\
		Evaluation  & & & & &  PMC & 3 \\
		(NBIDE) & & S\&W 9VE  & 4      &  1    &   Remington &  3 \\
		& & & & &  Winchester & 3 \\
		& & & & &  Speer & 3 \\
		& & & & &  PMC & 3 \\
		& & Sig Sauer P226  & 4      &  1    &   Remington &  3 \\
		& & & & &  Winchester & 3 \\
		& & & & &  Speer & 3 \\
		& & & & &  PMC & 3 \\
		\hline
		FBI: Colt & 90 & Various Colts & 45 & 1 & Remington & 2\\
		\hline
		FBI: Glock & 90 & Various Glocks & 45 & 1 & Remington & 2 \\		
		\hline
		FBI: Ruger & 100 & Various Rugers & 50 & 1 & Remington & 2\\
		\hline
		FBI: S\&W & 138 & Various S\&Ws & 69 & 1 & Remington & 2 \\
		\hline
		FBI: Sig Sauer & 130 & Various Sig Sauers & 65 & 1 & Remington & 2 \\
		\hline
		Todd Weller  &    95   &  Ruger P95DC   &  1   &     10 &     Winchester & 5-9 \\
		\hline
	\end{tabular}
\end{center}

For the 2D data, each casing was imaged using a Leica FS M reflectance microscope with ring light illumination. The objective was 2X, the resolution was 2.53 $\mu$m, and images are 1944 $\times$ 2592 pixel grayscale files in PNG format. Pixel values range from 0 to 255, where 0 represents black pixels and 255 represents white pixels. 3D data are measured using a Nanofocus $\mu$Surf disc scanning confocal microscope. Various magnifications were used, for example an objective of 10X results in a lateral resolution of $3.125 \mu m$, and images that are around $1200 \times 1200$. Pixel values are depth values in $\mu m$ (microns).

\section{Methodology}
\label{sec:method}

\subsection{General framework}
In order to automatically match 3D images of cartridge cases, we make use of a record linkage framework. Record linkage is the process of inferring which entries in different databases correspond to the same real-world identity, in the absence of a unique identifier \citep{Christen2012}. 
Beginning with records from two databases A and B to be linked, fields such as name and address are pre-processed and standardized. This results in some set of features for each record. Indexing methods might then be used to reduce the comparison space, for example candidate pairs in a demographic database might be restricted to those born in the same month, with all other pairs not being considered for linkage. Now pairwise comparisons are generated, consisting of one or more similarity measures. These pairwise data are then classified into matches and non-matches (and possibly a third category, potential matches). This classification step may be designed to incorporate restrictions such as one-to-one links, or preserving transitivity. A clerical review may also be conducted for potential matches, where a human examines pairs manually to decide the appropriate classification. Finally an evaluation of the results is conducted.

\begin{figure}[!ht]
	\centering
	\includegraphics[width = .8\textwidth]{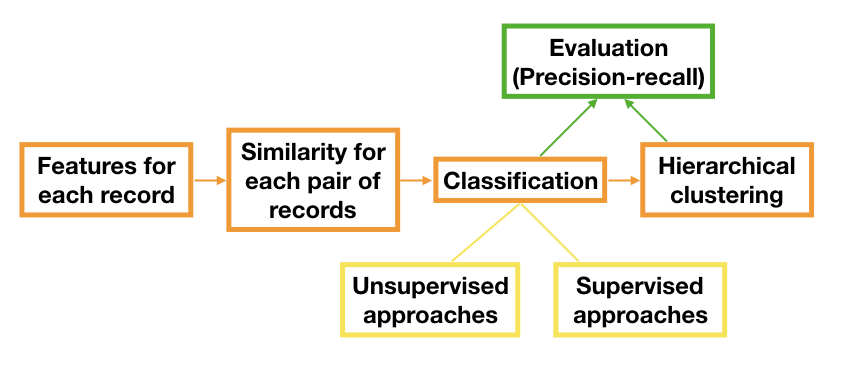}
	\caption{Simplified record linkage framework used to develop forensic matching methods. This is adapted from \cite{Tai:2018aa}.}
	\label{fig:RLframework}
\end{figure}

In \cite{Tai:2018aa}, we describe how forensic matching can be thought of as an application of record linkage; simply think of records as evidence samples, and real-world entity as the source of the sample. To develop methodology for matching 3D images, we use a simplified framework illustrated in Figure \ref{fig:RLframework}. Each of the steps is described in more detail in the following subsections.

\subsection{Features for each record}
At an individual record level, the data need to be pre-processed before useful features can be extracted. Pre-processing typically involves selecting and cleaning up the data fields. For cartridge cases, each record is an image, so the pre-processing steps are slightly more involved. The steps we use are 1) Automatically select breechface marks; 2) Level image; 3) Remove circular symmetry; 4) Filter. In \cite{Tai2018} we develop methodology for pre-processing 2D images; methodology for 3D images is adapted from the former. 
In Section \ref{sec:results} we compare the results using 2D and 3D data.

\textbf{\textit{Automatically select breechface marks}}

Looking at Figure \ref{fig:mechanism}\subref{fig:marks}, the area that we are interested in is the area with breechface marks. We would like to remove all other areas from the analysis, in particular the firing pin impression. The breechface region is relatively flat and has a different depth from the the firing pin impression, since the latter is an indentation on the primer surface. This is obvious from the rightmost image. The solution that we have adopted is to fit a plane through the breechface region, ignoring the firing pin impression, and to select only points lying on or close to the plane. 

We use an algorithm called RANdom SAmple Consensus (RANSAC) \citep{Fischler1981} to achieve this, and the procedure used is outlined in Algorithm \ref{alg:RANSAC}. RANSAC was developed to fit models in the presence of outliers. Briefly, we repeatedly fit planes to the image by sampling points on the image, and outliers are defined to be points lying outside some threshold from the fitted plane. The plane with the largest number of inliers (opposite of outliers) is chosen. These inliers are treated as being part of the breechface area.

\begin{algorithm}
	\caption{RANSAC to find best-fitting plane through Image $I$}
	\label{alg:RANSAC}
	\begin{algorithmic}[1] 
		\Procedure{bestPlane}{$I$} 
		\For{\texttt{i} in 1:\texttt{iter}} 
		\State Sample 3 points from $I$
		\State Fit plane to sampled points
		\State Count the number of inliers within preset threshold
		\EndFor 
		\State Select model with largest number of inliers
		\State Re-fit model using only inliers from selected model
		\State \textbf{return} fitted plane, selected inliers
		\EndProcedure
	\end{algorithmic}
\end{algorithm}

\begin{figure}[!ht]
	\centering
	\includegraphics[width = .7\textwidth]{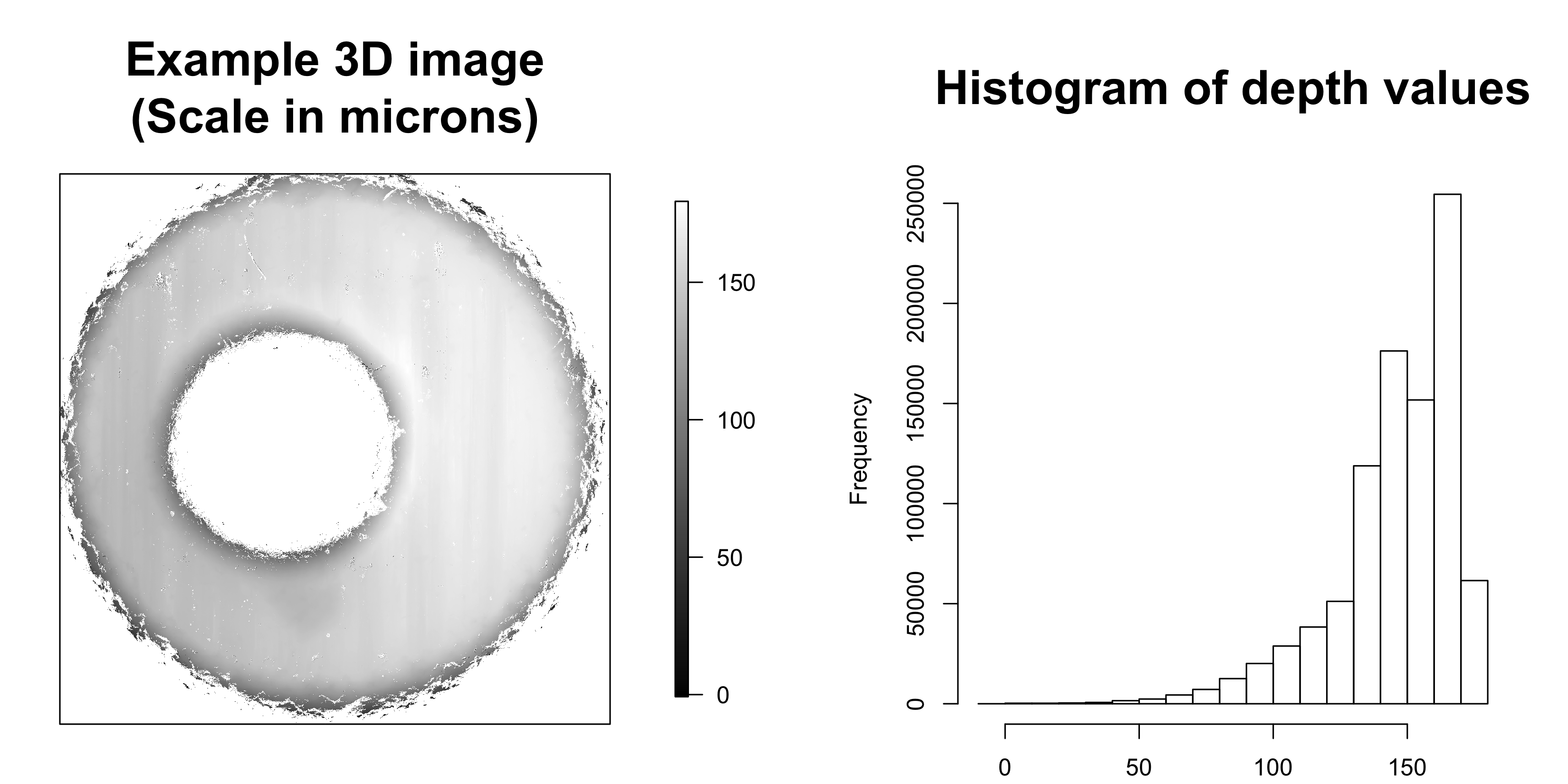}
	\caption{Typical 3D breechface image (without a firing pin impression captured) and associated histogram of depth values in microns. }	
	\label{fig:brerr}
\end{figure}

There are three parameters to be selected: $s$, the number of points sampled, $\delta$, the threshold above which points are considered to be outliers, and $N$, the number of iterations \texttt{iter} the algorithm runs for (in other words the number of samples). $s$ is typically the number of points required to fit the model; in this case 3 points are required to fit a plane. $\delta = 10\mu m$ is selected: Figure \ref{fig:brerr} shows a typical 3D breechface image (without a firing pin impression captured) and the associated histogram of depth values -- one can see that most of the depths are within $10\mu m$ of each other, making this a reasonable choice. 

As for $N$, it should be set to be large enough such that the probability of at least one random sample being free from outliers is at least $p$. Now, let $X$ be the random variable denoting the number of samples, out of $N$, having no outliers. Let $e$ be the proportion of outliers in the data, and $s$ be the number of sampled points. Then $X \sim \operatorname{Binom}(N, (1-e)^s)$, and the goal is to have $\mathbb{P}[X \geq 1] \geq p$.

Now, $ \mathbb{P}[X \geq 1] = 1 - \mathbb{P}[X = 0 ] = 1 - [1- (1-e)^s]^N$, so the objective becomes   
$$
\begin{aligned}
1 - [1- (1-e)^s]^N & \geq p.
\end{aligned}
$$
Taking logs, 
$$
\begin{aligned}
\log{(1 - p)} & \geq N \log{[1- (1-e)^s]} \\
N & \geq \frac{\log{(1-p)}}{\log{[1- (1-e)^s]}}
\end{aligned}
$$

\begin{figure}[!ht]
	
	\centering
		\includegraphics[width=\columnwidth]{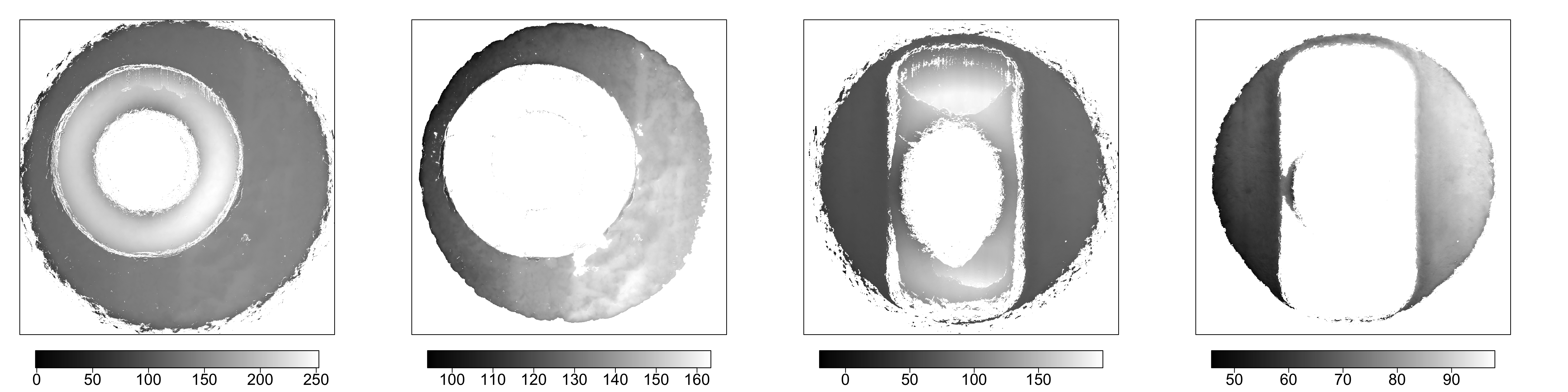}
\caption{\label{fig:ransacOutput}
	Examples of selected breechface areas after applying Algorithm \ref{alg:RANSAC}. The scale is in microns. The first set of images is of an example with a circular firing pin impression and its selected breechface marks, and the second is of an example with a rectangular firing pin impression.}

\end{figure}

Consider $p = .99$, $s = 3$ and $e = .6$. This last argument says that 60\% of points are outliers. This is set to be a relatively large proportion, since some images capture parts of the firing pin impression, and all these points would be considered to be outliers. Nevertheless, $e = .6$ is still likely to be an overestimate. This gives $N \geq 70$. In the actual implementation we use 75 iterations. Two examples of the resulting breechface impressions being selected are in Figure \ref{fig:ransacOutput}. Both circular and rectangular firing pin impressions can be removed well using Algorithm \ref{alg:RANSAC}. To illustrate further pre-processing steps, we use the first example in Figure \ref{fig:ransacOutput}.

\begin{figure}[ht]
	\centering
	\includegraphics[width = 5cm]{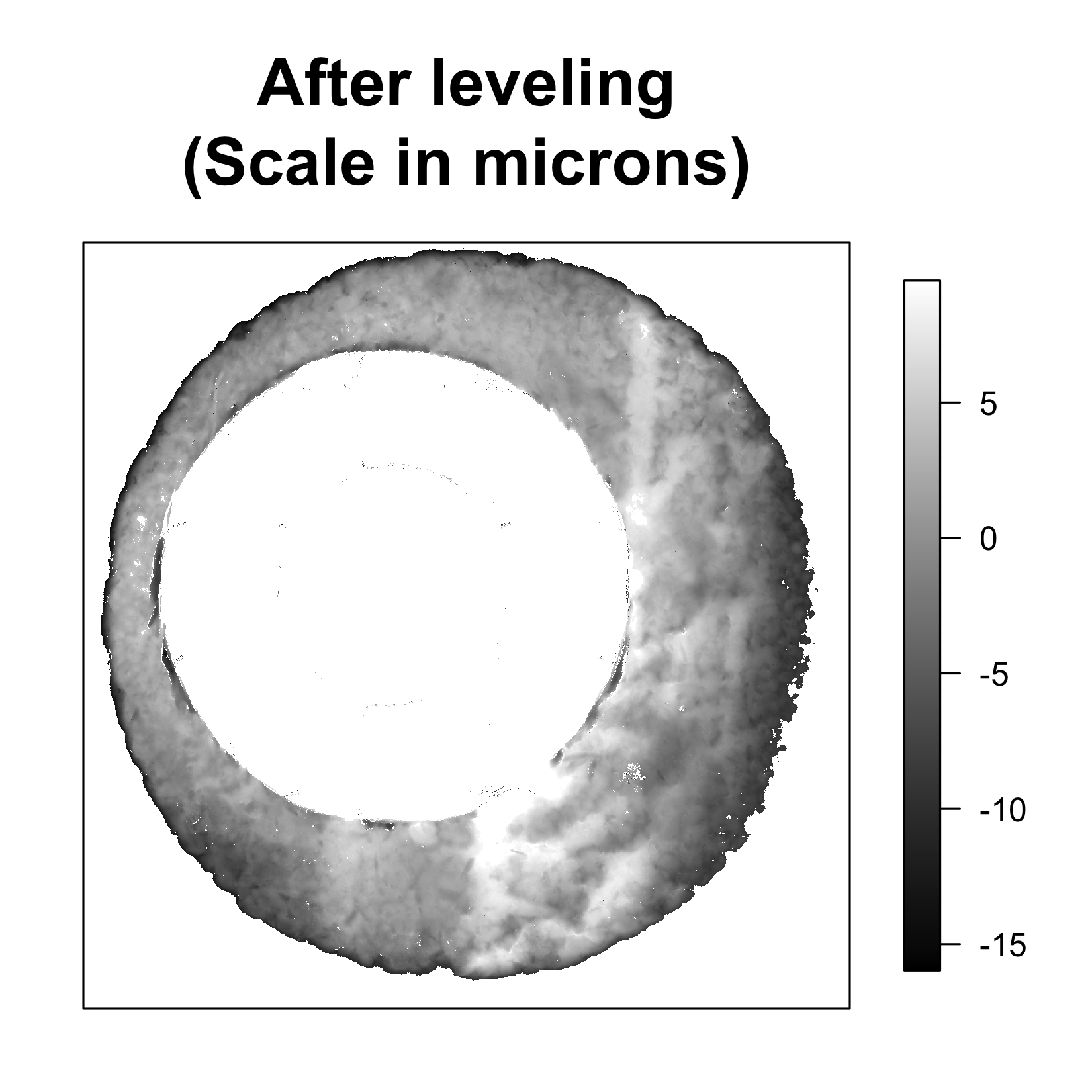}
	\caption{In Figure \ref{fig:ransacOutput}, the top-left of the first image is shallower than the bottom-right. This is resolved after leveling the image.}
	\label{fig:level}
\end{figure}

\textbf{\textit{Level image}}
This step adjusts for the breechface surface being tilted on a plane during the image capturing process, and has been used in the literature (see e.g., \cite{Vorburger2007}, \cite{Roth2015}). The same type of tilt in two images could result in high overall similarity, despite the two having no similar individual features. This is undesirable since one is less concerned with these overall patterns. To address this issue, we subtract the plane that was fit in the previous step, planar differences in depth. The residuals are used for further processing. The first example in Figure \ref{fig:ransacOutput}, after leveling, is in Figure \ref{fig:level}.

\textbf{\textit{Remove circular symmetry}}
Now, we standardize the different resolutions, by resizing all images to the lowest resolution of 6.25$\mu m$. It is at this resolution that all further pre-processing is done. The next step is to adjust for the surface having differences in depth that are circular in nature. This adapts the methodology of \cite{Tai2018}; here we show that the benefits transfer to topographical data, and provide the statistical details that were previously omitted. Analogous to the previous step, the base of the cartridge case could have differences in depth that are circular in nature. This could arise if the base of the cartridge case slopes inwards towards the center; looking at the second panel in Figure \ref{fig:mechanism}\subref{fig:marks}, this does look to be the case. The step of removing circular symmetry corrects for these circular differences in depth, by fitting a model that captures the differences. The residuals are used for further processing.

\begin{figure}[ht]
	\centering
	\includegraphics[width = 5cm]{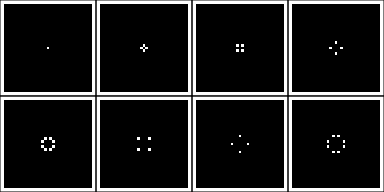}
	\caption{Illustration of the first eight matrices in a circularly symmetric basis. The locations in white are the same distance from the center, and matrices are enumerated from center outwards. 
	}
	\label{fig:basis}
\end{figure}

To model the circular differences in depth, we fit a linear combination of circularly symmetric basis functions \citep{Zeifman2017}. Consider a square matrix of dimension $m \times m$, where $m$ is odd. The center entry is at $\left( \frac{m+1}{2}, \frac{m+1}{2} \right)$, and such a matrix is said to be circularly symmetric if entries located the same distance from the center entry take the same value. Any matrix can be decomposed into a linear combination of the matrices in a circularly symmetric basis, plus residuals. The first few matrices in the basis are shown in Figure \ref{fig:basis}, where each panel represents one basis matrix. Each matrix takes the value 1 for pixels that are the same distance from the center, and zero otherwise. Bases are enumerated from center outwards.

To represent these matrices as basis functions, define functions $f_k$ for each matrix $k$, taking the $ij$-coordinates as inputs, and returning the value 0 or 1 depending on whether the input location is the required distance, $d_k$, from the center, i.e., if it is white or black in the pictorial representation in Figure \ref{fig:basis}. An example of a basis function (the fifth one, enumerated from the center outwards) is in Equation \ref{eq:basisFunction}. The decomposition of a matrix can then be represented as Equation \ref{eq:linearCombinationBasis}, where $K$ is the number of basis functions, $f_k$ is the $k$th basis function, $\beta_k$ is the basis function coefficient for $f_k$, and $\epsilon$ is the error.

\begin{equation}
\label{eq:basisFunction}
f_5(i,j)= \begin{cases} 1 &\mbox{if } d(i,j)=d_5 = \sqrt5 \\ 0 &\mbox{otherwise.} \end{cases} 
\end{equation}

\begin{equation}
\label{eq:linearCombinationBasis}
Image(i,j)=\sum_{k=1}^{K} \beta_k f_k(i,j)+\epsilon(i,j)
\end{equation}

The number of basis functions required, $K$, depends on the resolution of the image. For example, a $3 \times 3$ image has three possible distances from the center (including the corners), and can be represented using three basis functions. A $701 \times 701$ image (this is the size of the NIST 3D images after cropping and standardizing resolutions) requires 39,978 basis functions. The coefficient for each basis function is the mean of pixel values for pixels with the corresponding distance from the center. 

In this analysis missing pixels are ignored. Since only the breechface marks are selected for analysis, treating the other regions as missing pixels, some fraction of the coefficients will not be computed, but the total number of coefficients is still very large. There are large local fluctuations in the model coefficients since many of them are estimated only using a small number of pixels (many basis functions only have four pixels, and many of the pixels are also missing due to the selection of the breechface area). In order to only capture a global effect,  a loess regression \citep{Cleveland1979} is fit to the basis coefficients to reduce the variance of the estimates.

\begin{figure}[!ht]
	\centering
	\caption{Removing circular symmetry from the example image. \label{fig:circularResults}}
	\begin{subfigure}[t]{0.6\textwidth}
		\centering
		\includegraphics[width=\columnwidth]{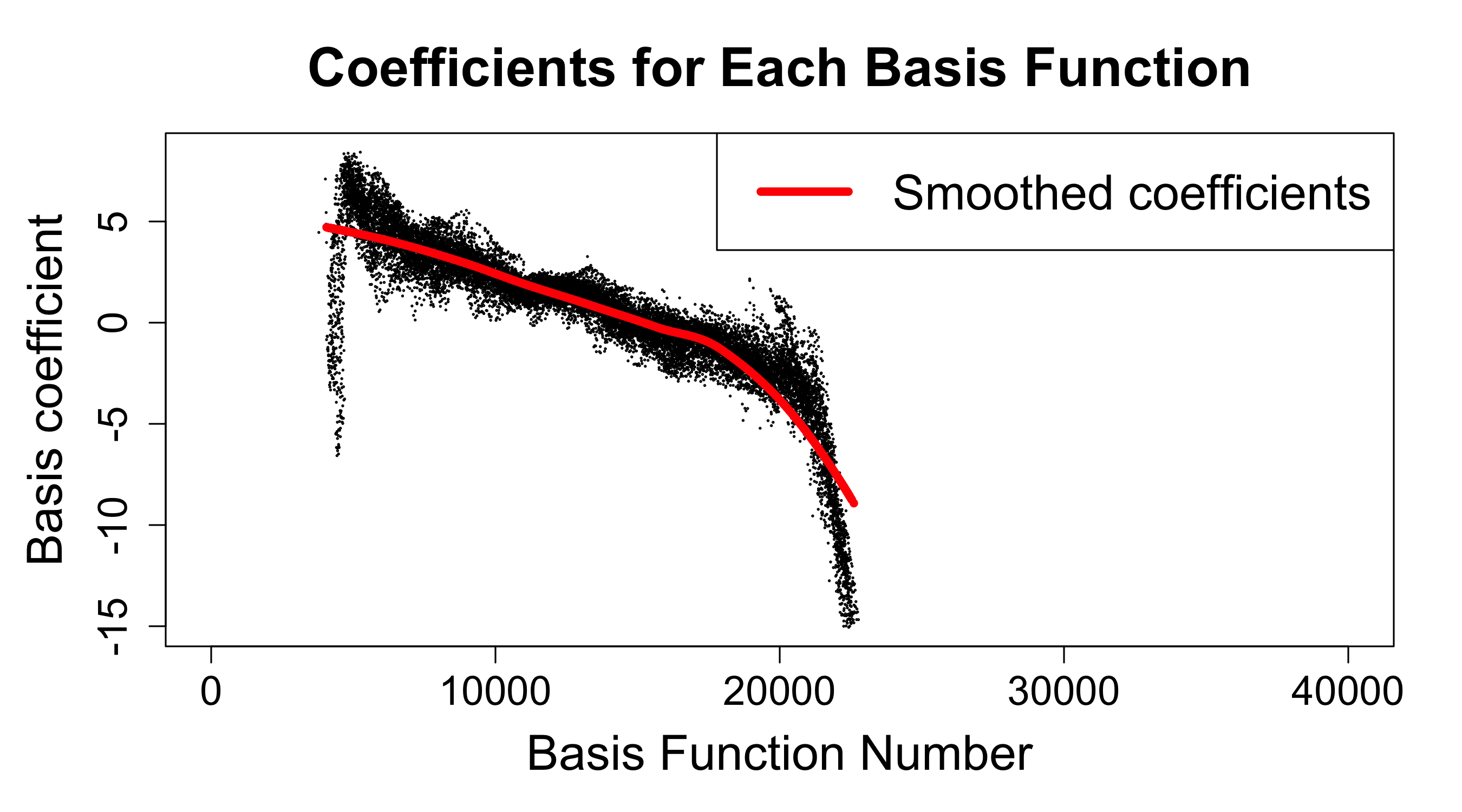}
		\caption{The set of circularly symmetric basis functions is fit to the example residual image after leveling (Figure \ref{fig:level}). \label{fig:basisCoefs}}
	\end{subfigure}
	~
	\begin{subfigure}[t]{0.35\textwidth}
		\centering
		\includegraphics[width = \columnwidth]{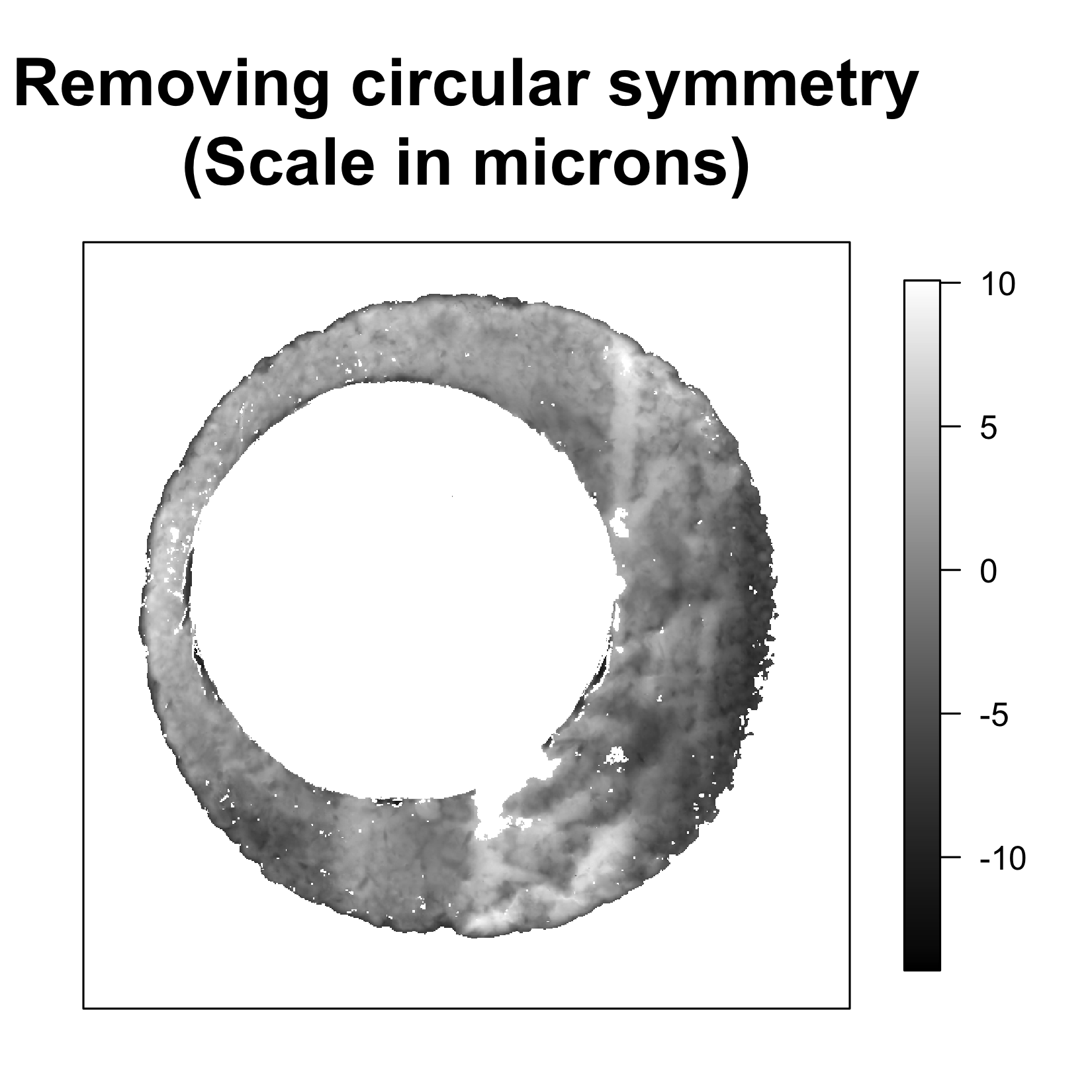}
		\caption{Residual image produced. \label{fig:removeCircular}}
	\end{subfigure}	

\end{figure}

Loess regression assumes that $y_i = f(x_i) + \epsilon_i$, where $\epsilon_i$ is the error. The function $f$ is approximated at each point in the range of the data set, using a low-degree polynomial and weighted least squares. Weights are inversely proportional to the distance from the point of interest. The parameters required for estimating the function are the degree of the polynomial, the weighting function, and the proportion of all points to be included in the fit. Here a quadratic function is fit, and the proportion of points to be included in the fit is set to be $\alpha = 0.75$. The weight function (for local points that are included) from the point $x_0$ is computed using a tricubic kernel $w_i(x_0) = \left(1 - \frac{|x_i - x_0|}{h_0}^3\right)^3$, where $h_0$ is the $(\alpha *n)^{\text{th}}$ largest $|x_i - x_0|$. A large value of $\alpha$ was selected for a large degree of smoothing, as can be seen in Figure \ref{fig:circularResults}\subref{fig:basisCoefs}. With this large degree of smoothing, the loess regression is not particularly sensitive to the weight function or degree of polynomial, and we simply used the default values implemented in the \texttt{loess()} function in \texttt{R}. 

Loess regression is a linear smoother, meaning that the fitted values are a linear combination of the original values \citep{Buja:1989aa}. Precisely, $\mathbf{\hat{Y}} = \mathbf{SY}$, where $\mathbf{S}$ is an $n\times n$ matrix, called a smoother matrix. The smoother matrix is analogous to the hat matrix in multiple linear regression: $\mathbf{\hat{Y}} = \mathbf{HY}$, where $\mathbf{H} = \mathbf{X(X^TX)^{-1}X^TY}$. The trace of the smoother matrix is the effective degrees of freedom of the loess regression, and using these specifications the effective degrees of freedom is less than 5 for each image. Results for the same example image are in Figure \ref{fig:circularResults}. There is evidence of an inward slope towards the center, which we hypothesized based on Figure \ref{fig:mechanism}\subref{fig:marks}. 

The fitted model and the residuals for the same example image are in Figure \ref{fig:circularResults}. The two identified sources of bias, planar and circular, are no longer present in the residuals.

\textbf{\textit{Filter}}
The last pre-processing step is filtering, which again has been used in the literature. 
Filtering highlights certain features of the image. A Gaussian filter is used, with short and long cutoffs approximately 20-150$\mu m$, meaning that patterns with wavelengths within this range are highlighted. Empirically these produced the best results among some test examples; as far as we know there is no established guidance on what wavelengths are relevant for the comparison. Some cutoffs that have been used in other work are 2.5-250$\mu m$, 37-150$\mu m$ \citep{Vorburger2007}, and 16-250$\mu m$ \citep{Song2018}.

The resulting image after all pre-processing is on the left in Figure \ref{fig:comparison}. Note the prominence of marks compared to the original image in the left panel of Figure \ref{fig:ransacOutput}. 




\subsection{Pairwise comparison: Similarity for each pair of records}
Given features for individual records, the next step is to generate meaningful similarity scores for pairwise comparisons. In this particular application, it is necessary to first align the two images to one another. Aligning images typically involves finding the best rotation and translation parameters (horizontal and vertical), where ``best'' means some metric is optimized, over all pixels or some subset of pixels. This optimization can be done either using a grid search or some other approximate means. The metric used might be correlation, mean-squared error, or something similar. Frequently, the optimized value of the metric is used as a similarity measure; this might be used on its own or combined with other similarity measures to produce a feature set for comparing two images. 

In the literature, perhaps the most widely used measure of similarity is the maximized correlation over rotations and translations, known as $CCF_{max}$, or maximum cross-correlation function \citep[see e.g.,][]{Vorburger2007,Roth2015,Riva2014,Geradts2001}. This is what we use here. To be precise, the cross-correlation between two zero-mean images $I_1$ and $I_2$ (or more generally, 2-dimensional matrices), is defined as 
\begin{equation} \label{eq:ccf}
CCF_{I_1, I_2}(k, l) = \frac{\sum_{i, j} I_1(i, j)I_2(i + k, j + l)}{\sqrt{\sum_{i, j} I_1(i, j)^2} \sqrt{\sum_{i, j} I_2(i, j)^2}},
\end{equation}
where $(k,l)$ represents spatial lag (translation; with $k$ and $l$ being the vertical and horizontal lags respectively), $i$ indexes the rows and $j$ indexes the columns. The computation is repeated for different rotation angles, and $CCF_{max}$ is then the maximum correlation, taking into account rotations and translations.

We use a grid search to maximize the correlation between two images over translations and rotations. The maximized correlation derived in this manner is an estimator of $CCF_{max}$, or more precisely, $CCF_{I_1, I_2}^{max}$, where the two images being compared are $I_1$ and $I_2$. The details are in Algorithm \ref{alg:align2to1}.

\begin{algorithm}[!ht]
	\caption{Aligning image 2 to image 1}
	\label{alg:align2to1}
	\begin{algorithmic}[1] 
		\Procedure{align}{$I_1,I_2$} 
		\State \texttt{thetas} $ \gets -175, -170, \ldots, -10, -7.5, -5, \ldots, 5, 7.5, 10, 15, \ldots, 180$
		\State Scale $I_1$  (so $\bar{I_1} = 0$ and $\sum_{ij} I_1(i, j)^2 = 1$) \Comment{So that maximizing correlation $\equiv$ minimizing MSE}
		\For{\texttt{theta} in \texttt{thetas}} 
		\State Rotate $I_2$ by \texttt{theta} \label{lst:line:1}
		\State Scale rotated $I_2$ 
		\State Compute cross-correlation, $CCF_{I_1, I_2}^{\theta}(k, l)$, as in Equation \ref{eq:ccf}, for integer $k$ and $l$
		\State $CCF_{I_1, I_2}^{\theta} \gets \max_{k,l} CCF_{I_1, I_2}(k, l) $ \label{lst:line:3}
		\EndFor 
		\State $\theta'  \gets \arg \max_{\theta} CCF_{I_1, I_2}^{\theta} $ \Comment{Neighborhood of best $\theta$}
		\If{$\theta' \in [-10, 10]$}
		\State \texttt{fineThetas} $ \gets \theta' - 2, \theta' - 1.5, \ldots \theta' + 2$
		\Else 
		\State \texttt{fineThetas} $ \gets \theta' - 4, \theta' - 3, \ldots \theta' + 4$
		\EndIf
		\For{\texttt{theta} in \texttt{fineThetas}}  \Comment{Finer search in neighborhood of $\theta'$}
		\State Lines \ref{lst:line:1}-\ref{lst:line:3}
		\EndFor 
		\State $\theta^\star, k_{\theta}^\star, l_{\theta}^\star  \gets \arg \max_{\theta'} CCF_{I_1, I_2}^{\theta'} $            
		\State $CCF_{I_1, I_2}^{max}  \gets \max_{\theta'} CCF_{I_1, I_2}^{\theta'} $            
		\State \textbf{return} $CCF_{I_1, I_2}^{max}, \theta^\star, k_{\theta}^\star, l_{\theta}^\star$ 
		\EndProcedure
	\end{algorithmic}
\end{algorithm}

Let the correlation returned by Algorithm \ref{alg:align2to1} be $\hat{c_{12}}$. Now, for a comparison of $I_1$ and $I_2$, one can either align $I_2$ to $I_1$ using  \textsc{align}($I_1$, $I_2$) to return $\hat{c_{12}}$, or $I_1$ to $I_2$ using \textsc{align}($I_2$, $I_1$) to return $\hat{c_{21}}$. This might give slightly different results due to computational issues such as interpolation that is involved during a rotation. We compare $\hat{c_{12}}$ to $\hat{c_{21}}$ for each of the data sets, and note that the differences are minimal. We define the similarity score between two images to be 

\begin{equation}
\label{eq:simABBA}
\begin{split}
\hat{s_{12}} & = \max (\hat{c_{12}}, \hat{c_{21}}) \\
& = \max ( \textsc{align}(I_1, I_2), \textsc{align}(I_2, I_1) ),
\end{split}
\end{equation}

and this single pairwise feature is used for classification.

\begin{figure}[!ht]
	\centering
	\includegraphics[width = 8cm]{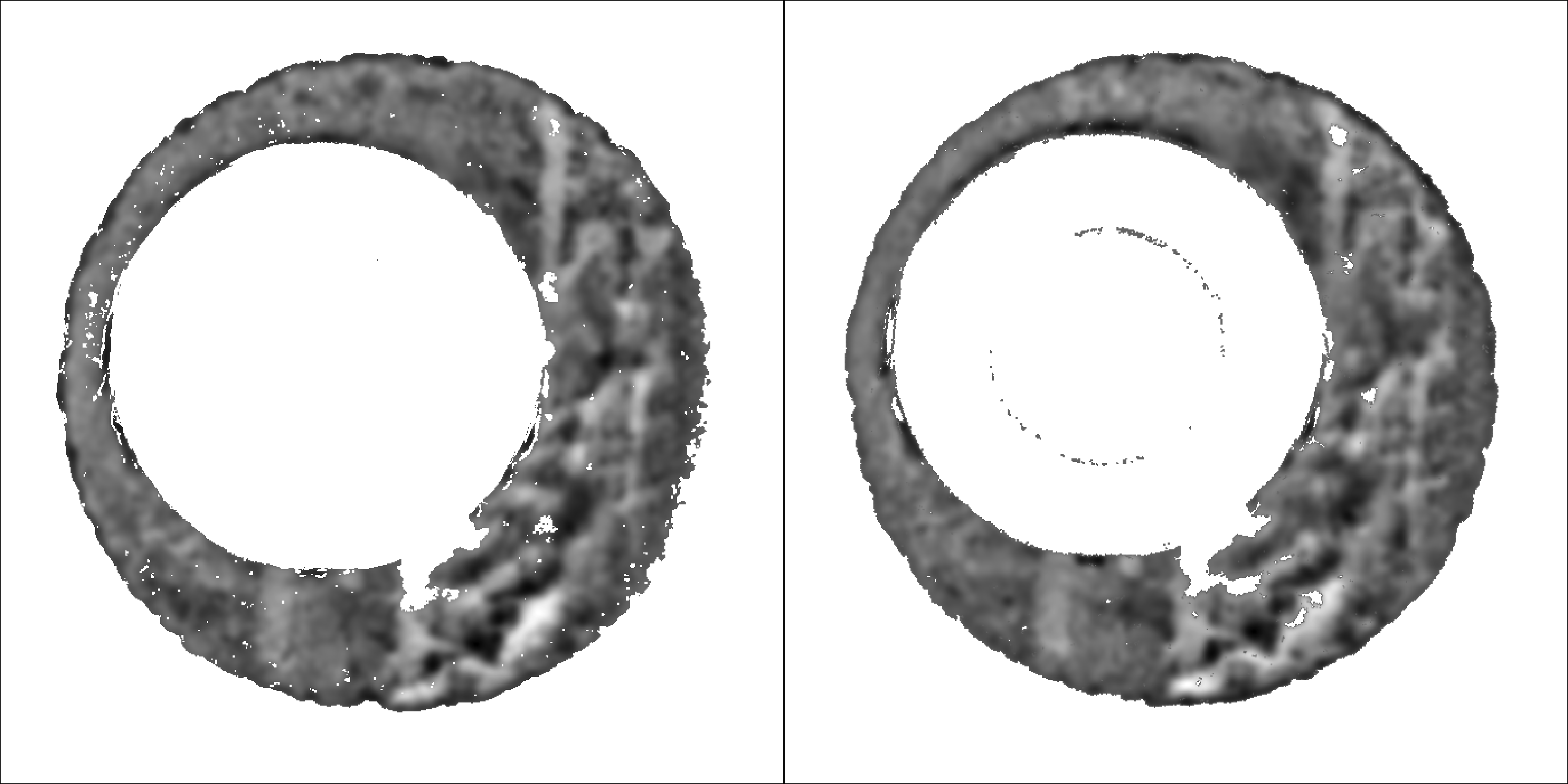}
	\caption{The example image from the previous figures is on the left. On the right is a processed image from the same gun, after alignment. Note that the parallel marks are now lined up well.}
	\label{fig:comparison}
\end{figure}

To illustrate the results, Algorithm \ref{alg:align2to1} is applied to a pair of example images in Figure \ref{fig:comparison}, where both are from the same gun. A similarity score of $0.76$ is obtained, with a rotation angle of $13^\circ$, meaning that the second image is rotated $13^\circ$ clockwise for best alignment. Plotting the two images with the second rotated, notice that the breechface marks are now lined up well. 


\subsection{Classification}
After pairwise similarity scores have been generated, there are numerous methods to classify the pairs into predicted matches and non-matches. We use a threshold-based approach, meaning that cutoffs on $\hat{s}$ can be used, above which pairs are classified as matches. The effects of choosing various cutoffs are examined using precision-recall graphs (see \ref{ssec:evaluation}).

\subsection{Hierarchical clustering}
In record linkage, additional restrictions are sometimes imposed in the matching process. For example, a one-to-one constraint may be added, meaning that a record in database A can be matched to only one record in database B. Another common situation is when multiple pairwise comparisons are done independently, and there is a desire to eliminate intransitive links. For example, if A matches B and B matches C, but A does not match C, there is a conflict that needs to be resolved. Here we use hierarchical clustering to eliminate intransitive links, in situations where these intransitive links are undesirable. For example, we might be interested in comparing results of an automatic method with that of examiner proficiency tests. In these tests multiple samples from test fires using the reference firearm are provided as reference samples. Examiners need to determine if various questioned samples come from the same source as the reference samples. Essentially, this involves organizing samples into groups that examiners believe come from the same source. These are necessarily transitive in terms of pairwise comparisons, and is arguably a simpler task than getting each pairwise comparison correct. To compare these results accurately with results from an automated method would require that the latter also have transitive links. Additionally, reference data sets are often used in developing automated methods. Researchers perform all pairwise comparisons within these data sets, and if there are intransitive links there would be inconsistencies in the evaluation. A final argument for imposing the constraint of transitivity is that it can lead to better classifier performance.

As far as we know, the issue of intransitivity has not been seriously considered in the forensic matching literature. It is possible that it is not a concern depending on the goals of the analysis, but we argue for its inclusion as an optional post-processing step. We adopt the approach in \cite{Ventura2015}, where given pairwise similarity predictions $s$, we use $d = 1 - s$ as a distance measure, and perform hierarchical agglomerative clustering using various linkage methods. 

Briefly, given items $1, \ldots, n$, dissimilarities $d_{ij}$ between each pair $i$ and $j$, and dissimilarities $d(G, H)$ between groups $G = \{i_1, i_2, \ldots i_r\}$ and $H = \{j_1, j_2, \ldots, j_s\}$, hierarchical agglomerative clustering starts with each node in a single group, and repeatedly merges groups such that $d(G, H)$ is within a threshold $D$. $d(G, H)$ is determined by the linkage method, and here we use single, complete, average \citep{Everitt:2001aa} and minimax linkage \citep{Bien2011}. These linkage methods define $d(G, H)$, as follows.

\begin{itemize}
	\item[] \textbf{Single linkage} $d_{\text{single}}(G, H) = \min_{i \in G, j \in H} d_{ij}$
	\item[] \textbf{Complete linkage} $d_{\text{complete}}(G, H) = \max_{i \in G, j \in H} d_{ij}$
	\item[] \textbf{Average linkage} $d_{\text{average}}(G, H) = \frac{1}{|G||H|} \sum_{i \in G, j \in H} d_{ij}$
	\item[] \textbf{Minimax linkage} $d_{\text{minimax}}(G, H) = \min_{i \in G\cup H} r(i, G \cup H)$, where $r(i, G) = \max_{j \in G} d_{ij}$, the radius of a group of nodes $G$ around $i$
\end{itemize}

Employing a linkage method and making a choice of cutoff would result in a data set that is grouped into clusters that share the same source. 

%
%

\subsection{Evaluation} \label{ssec:evaluation}
The classification step generates predictions for whether pairs are matches or non-matches (depending on some cutoff). Similarly, after hierarchical clustering the comparisons can be evaluated on a pairwise level. Since the metadata provided in NIST's database gives information on the source of each image, ground truth labels for whether pairs are from the same gun (a match or non-match) can be generated. One can use the same performance measures used in classification problems, at a pairwise level. 

In the forensics literature, evaluation on reference data is done in a haphazard manner. Most commonly, plots of score distributions from true matched and non-matched pairs are displayed as histograms \citep{Vorburger2007,Ott2017,Song2018,Roth2015,Riva2014,Lilien2017}, and a comment is made based on the visual separation between the distributions. If there is good separation between scores for matched and non-matched pairs, the method is said to be successful. Other work estimates the overlapping area between distributions, or the number of true matches in among the top 10 pairs in terms of similarity scores \citep{Vorburger2007}. Some other examples are the mean and standard deviation of the distributions of scores for matched and non-matched pairs \citep{Ott2017}, false positive and false negative rates, false discovery and false omission rates \citep{Song2018}, and ROC curves \citep{Roth2015}.



We propose standardizing the evaluation process by reporting precision and recall, as well as the area under the precision-recall graph as a numerical summary. This gives a better characterization of classifier performance, as opposed to visually comparing the two distributions. It is also an improvement over estimating the overlap region, since the latter often requires assumptions to be made about the data. It is more informative than other summary measures like mean and standard deviation of the distributions.

The choice of precision and recall is explained as follows. A typical confusion matrix in a classification problem has the following cells: 
%
%

\begin{center}
	\begin{tabular}{lcc}
		\multicolumn{3}{c}{Predicted} \\
		\cline{2-3}
		Truth & 0 & 1 \\
		\hline
		0 & True Negative & False Positive \\
		1 & False Negative  & True Positive \\
		\hline
	\end{tabular}
\end{center}

In record linkage problems, there is a large class imbalance since most pairs do not match, so typically, pairs are overwhelmingly true negatives. Measures that have the number of true negatives in the denominator, such as false positive rate, will often be close to zero. Hence we propose using $\text{Precision} = \frac{\text{True Positives}}{\text{True Positives + False Positives}}$ and $\text{Recall} = \frac{\text{True Positives}}{\text{True Positives + False Negatives}}$. Both measures range from 0 to 1, with 1 being perfect classifier performance. 

Since in our case the cutoff for predicting positives (matches) is not necessarily fixed (more details in Section \ref{sssec:finalClusters}), we compute the area under the precision-recall graph. This also ranges from 0 to 1, and maximum performance is achieved when the graph passes through the top-right corner, giving an area under the curve of 1.


\section{Results}
\label{sec:results}

We first present results of our proposed methodology in Section \ref{ssec:3D}, followed by comparisons between 3D and 2D data.

\subsection{3D results} \label{ssec:3D}

\subsubsection{Classifier and clustering accuracy}

\begin{figure}[!ht]
	\caption{\label{fig:3DbyStudy}
		Distribution of $\hat{s}$ for matches and non-matches by data set using 3D topographies. }
	
	\centering
	\begin{subfigure}[t]{0.3\textwidth}
		\centering
		\includegraphics[width=\columnwidth]{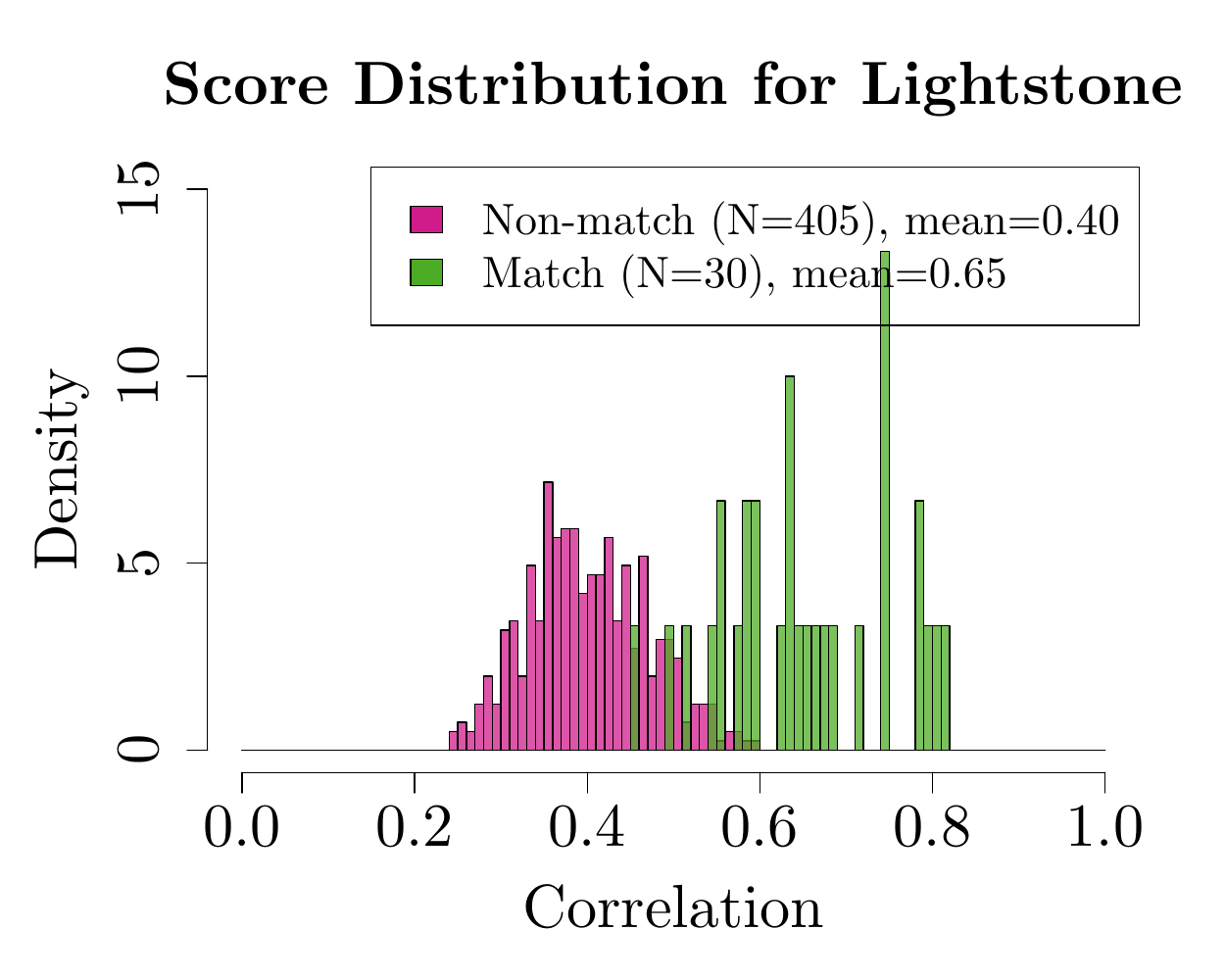}
		\caption{\label{fig:lightstone3D} Lightstone (consecutively manufactured)}
	\end{subfigure}
	~
	\begin{subfigure}[t]{0.3\textwidth}
		\centering
		\includegraphics[width=\columnwidth]{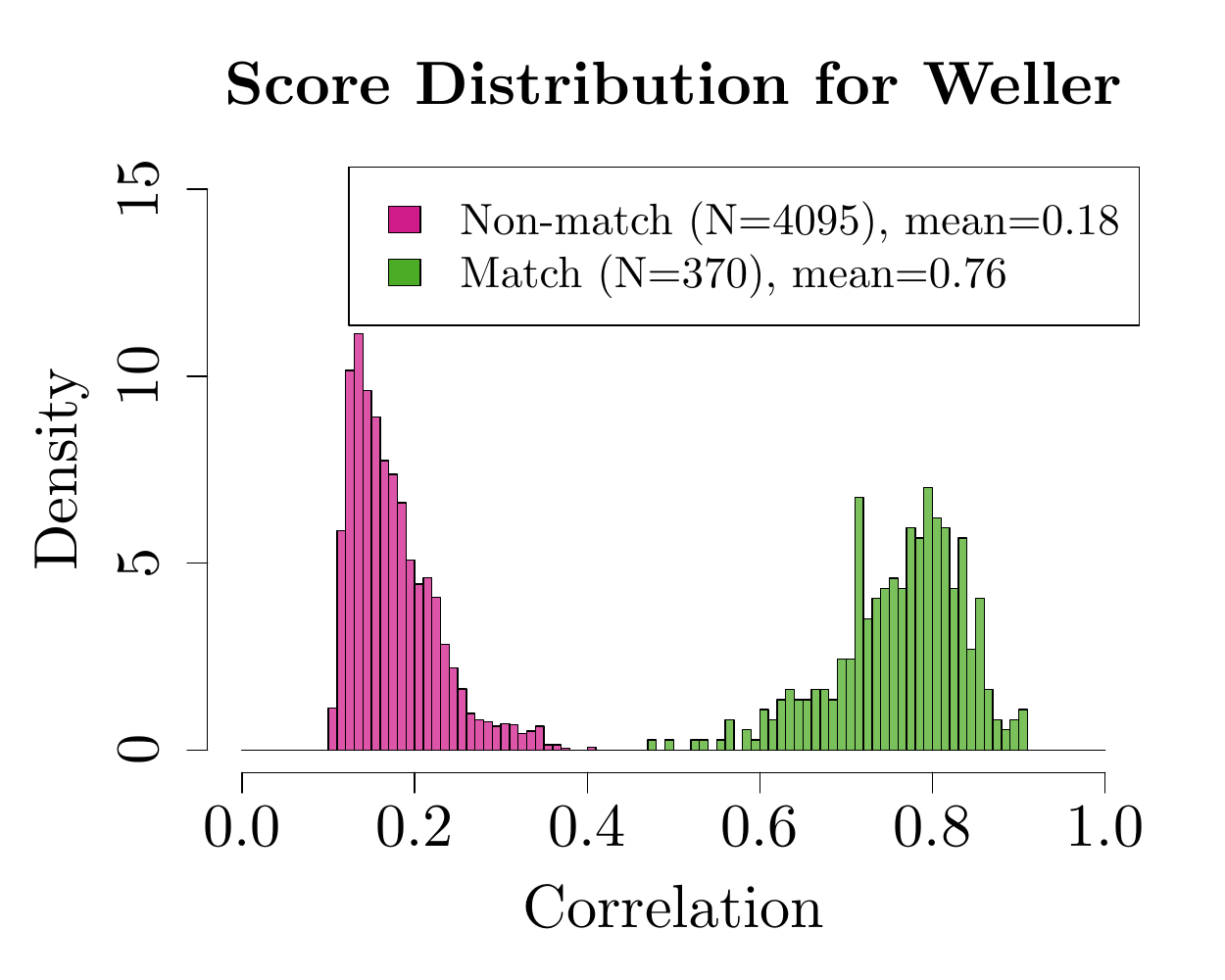}
		\caption{\label{fig:weller3D} Weller (consecutively manufactured)}
	\end{subfigure}
	~
	\begin{subfigure}[t]{0.3\textwidth}
		\centering
		\includegraphics[width=\columnwidth]{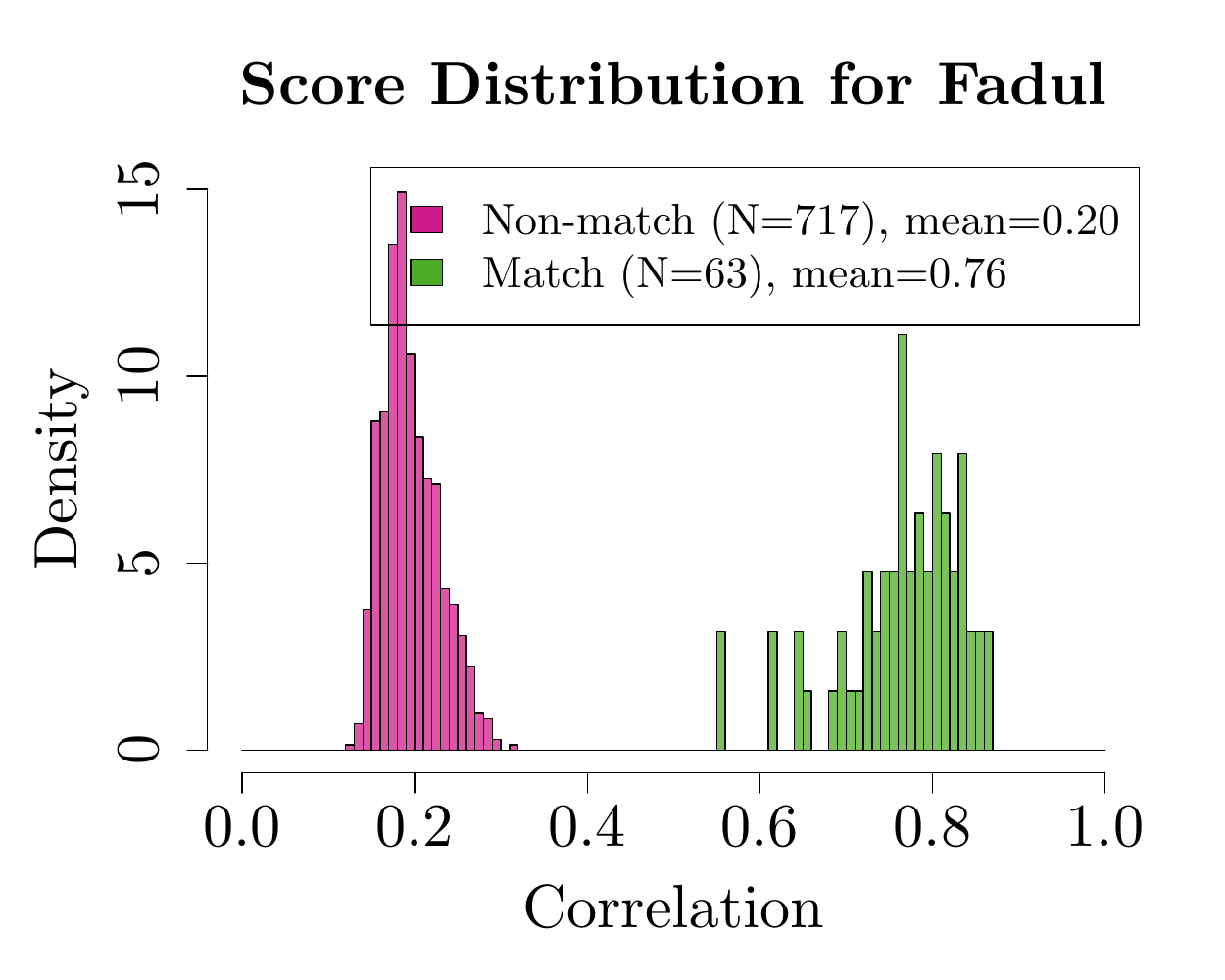}
		\caption{\label{fig:fadul3D} Fadul (consecutively manufactured)}
	\end{subfigure}
	~
	\begin{subfigure}[t]{0.3\textwidth}
		\centering
		\includegraphics[width=\columnwidth]{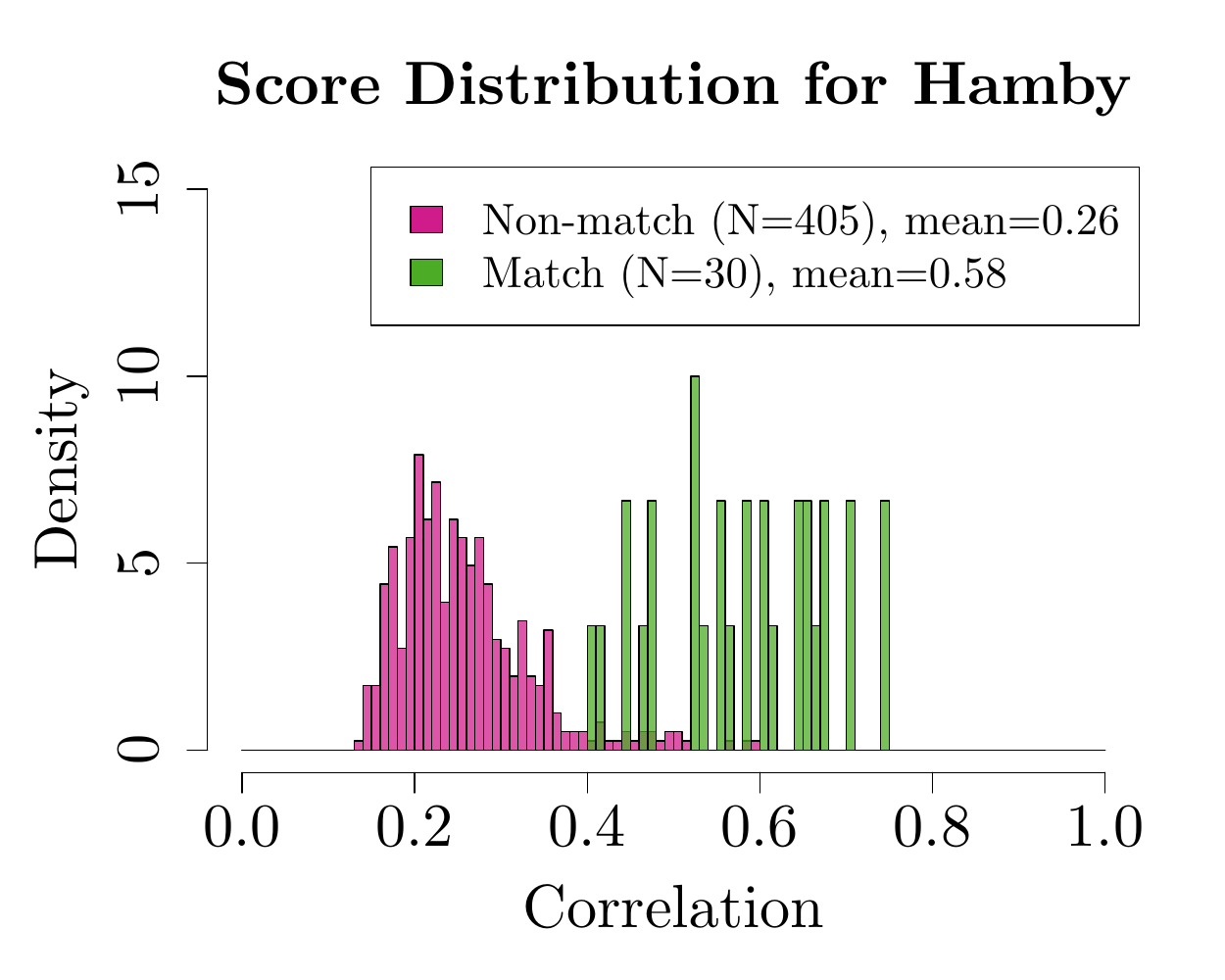}
		\caption{\label{fig:hamby3D} Hamby (consecutively manufactured)}
	\end{subfigure}
	~
	\begin{subfigure}[t]{0.3\textwidth}
		\centering
		\includegraphics[width=\columnwidth]{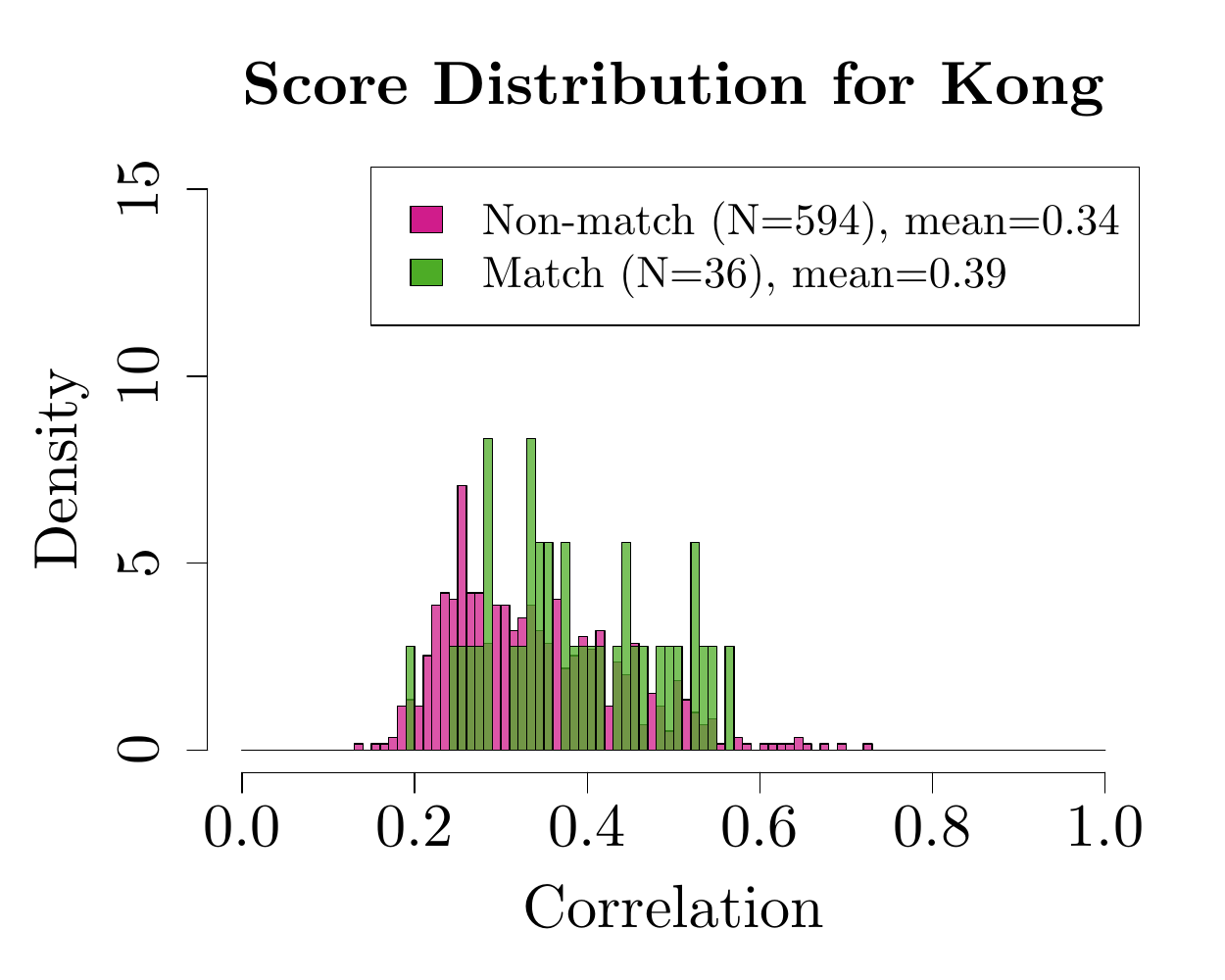}
		\caption{\label{fig:kong3D} Kong}
	\end{subfigure}
	~
	\begin{subfigure}[t]{0.3\textwidth}
		\centering
		\includegraphics[width=\columnwidth]{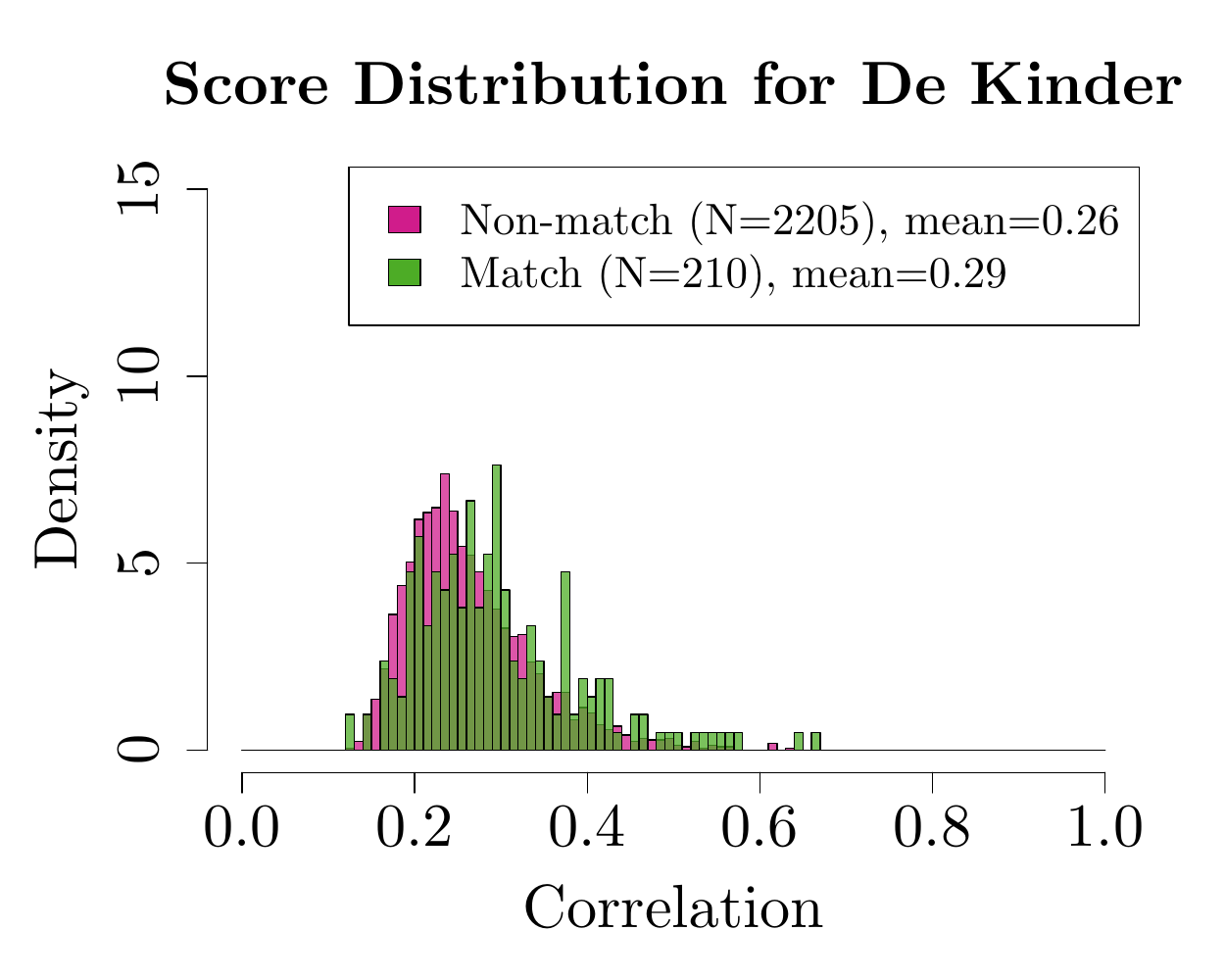}
		\caption{\label{fig:dekinder3D} De Kinder}
	\end{subfigure}
	~
	\begin{subfigure}[t]{0.3\textwidth}
		\centering
		\includegraphics[width=\columnwidth]{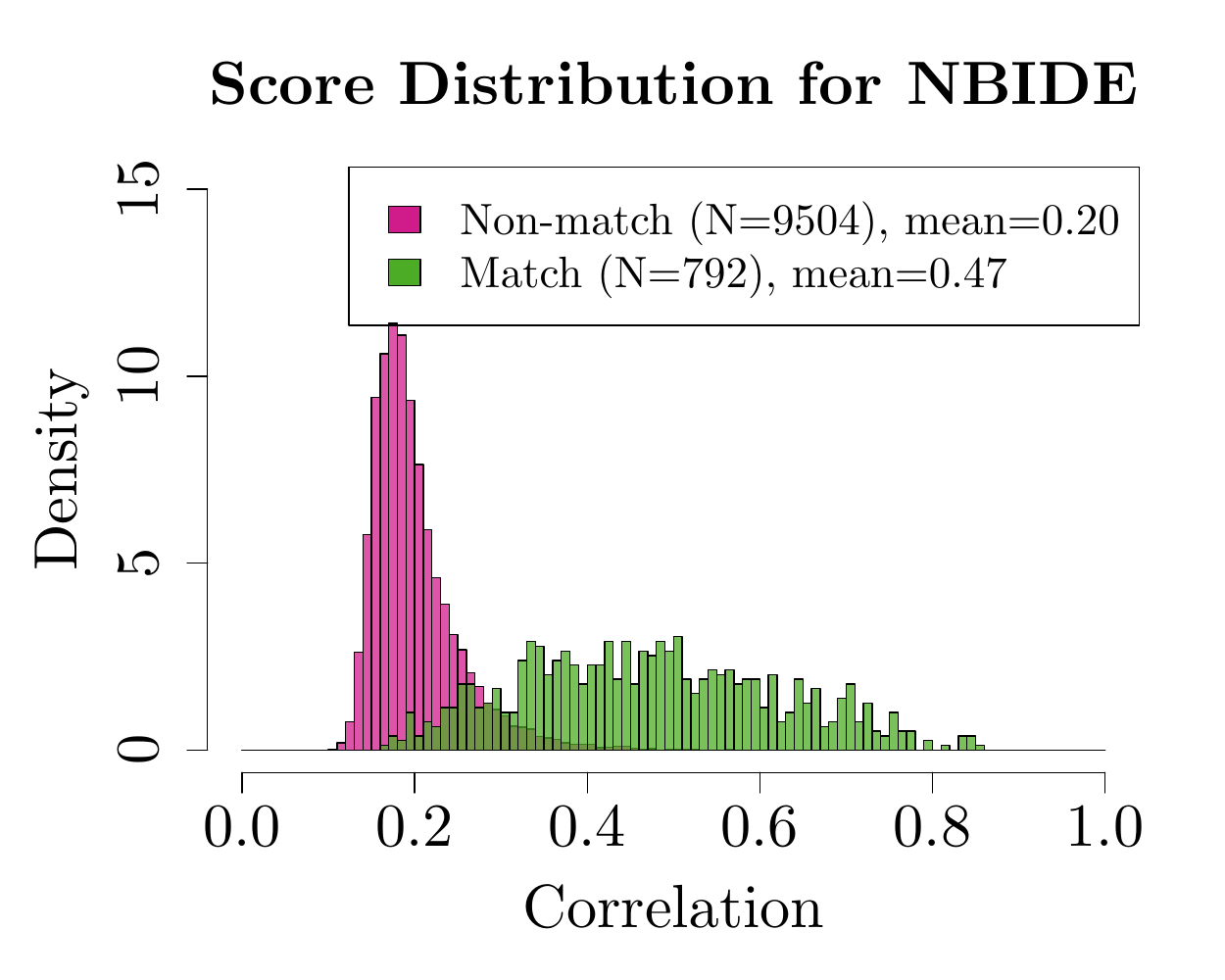}
		\caption{\label{fig:NBIDE3D} NBIDE}
	\end{subfigure}
	~
	\begin{subfigure}[t]{0.3\textwidth}
		\centering
		\includegraphics[width=\columnwidth]{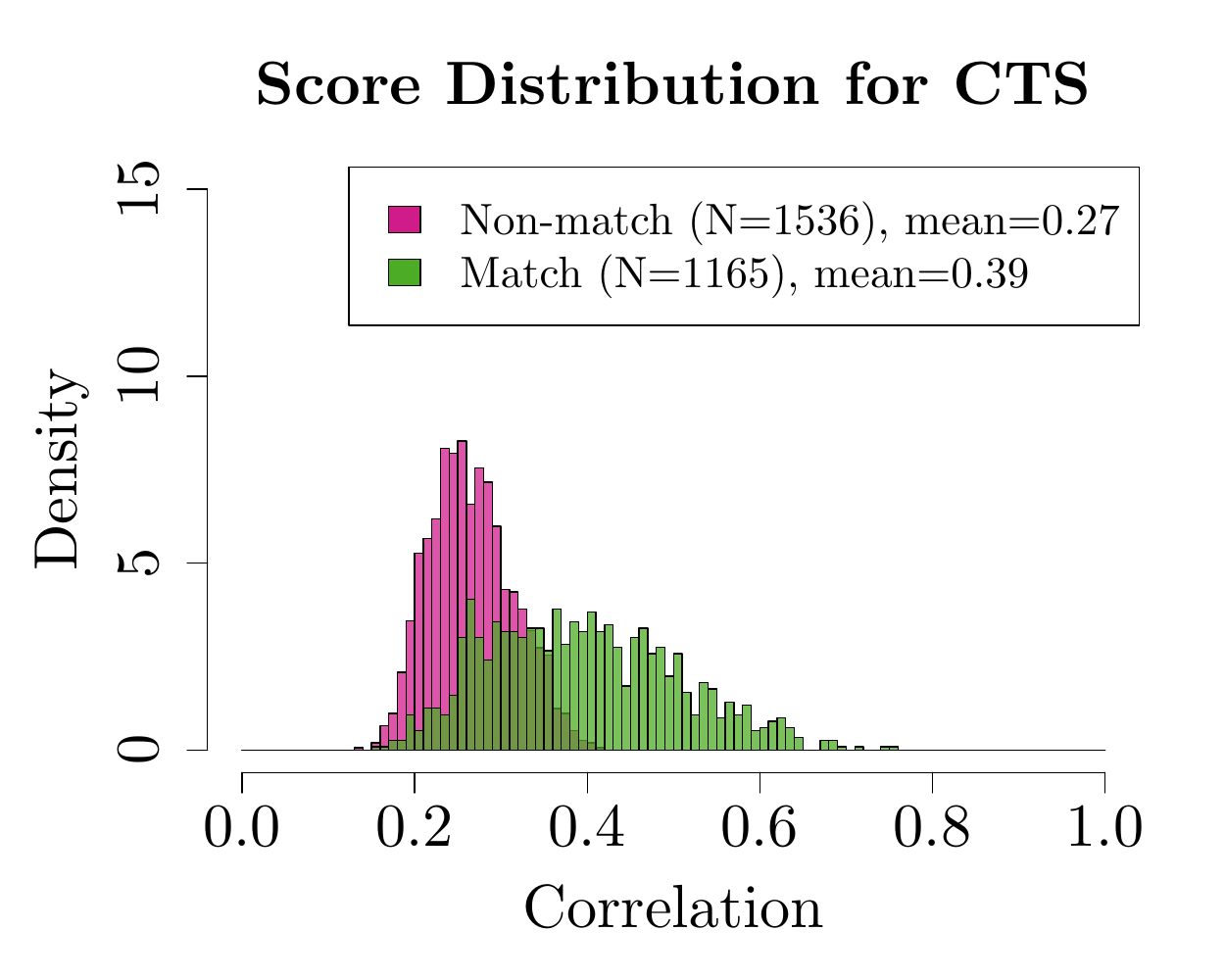}
		\caption{\label{fig:CTS3D} CTS}
	\end{subfigure}
	~
	\begin{subfigure}[t]{0.3\textwidth}
		\centering
		\includegraphics[width=\columnwidth]{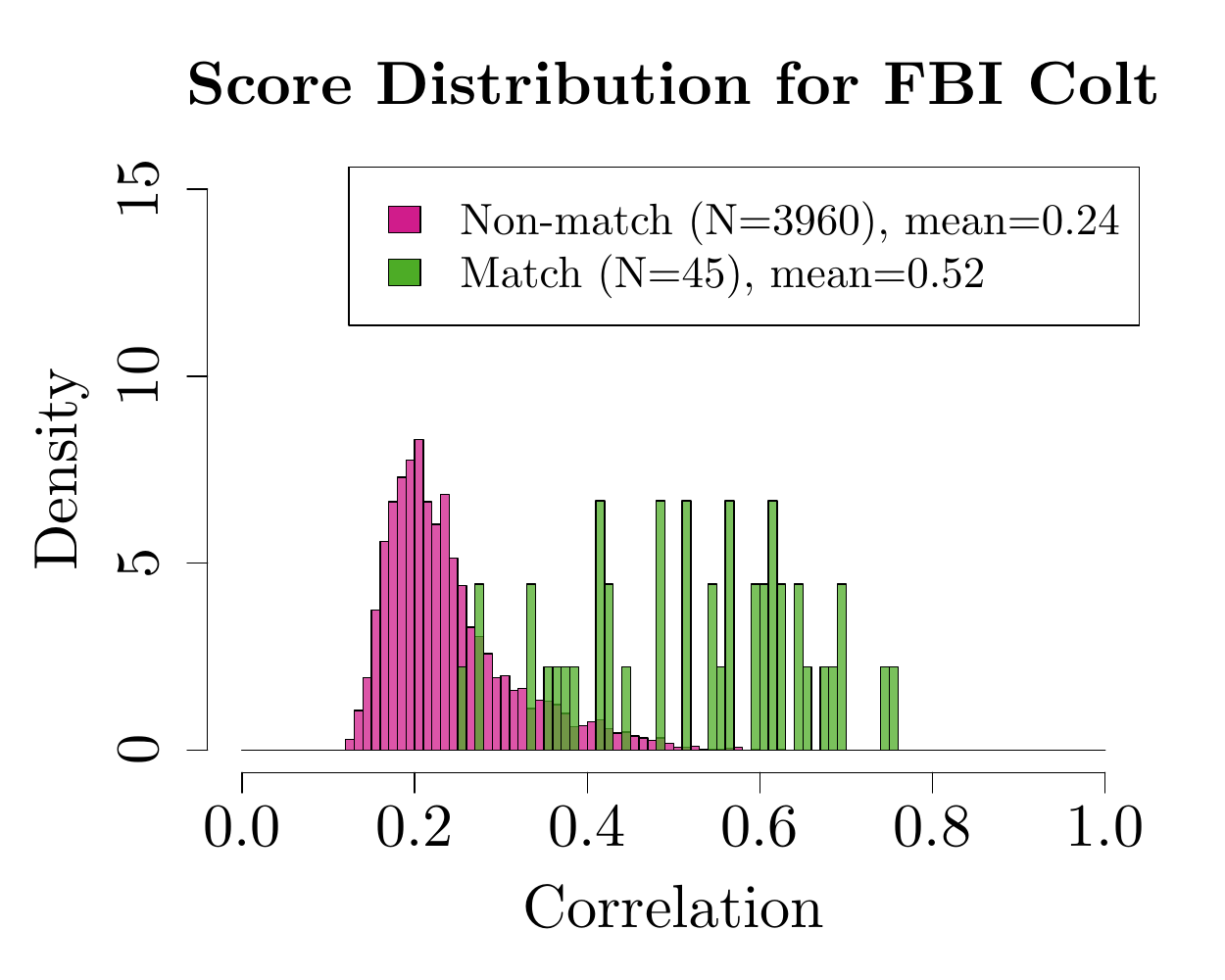}
		\caption{\label{fig:colt3D} FBI Colt}
	\end{subfigure}
	~
	\begin{subfigure}[t]{0.3\textwidth}
		\centering
		\includegraphics[width=\columnwidth]{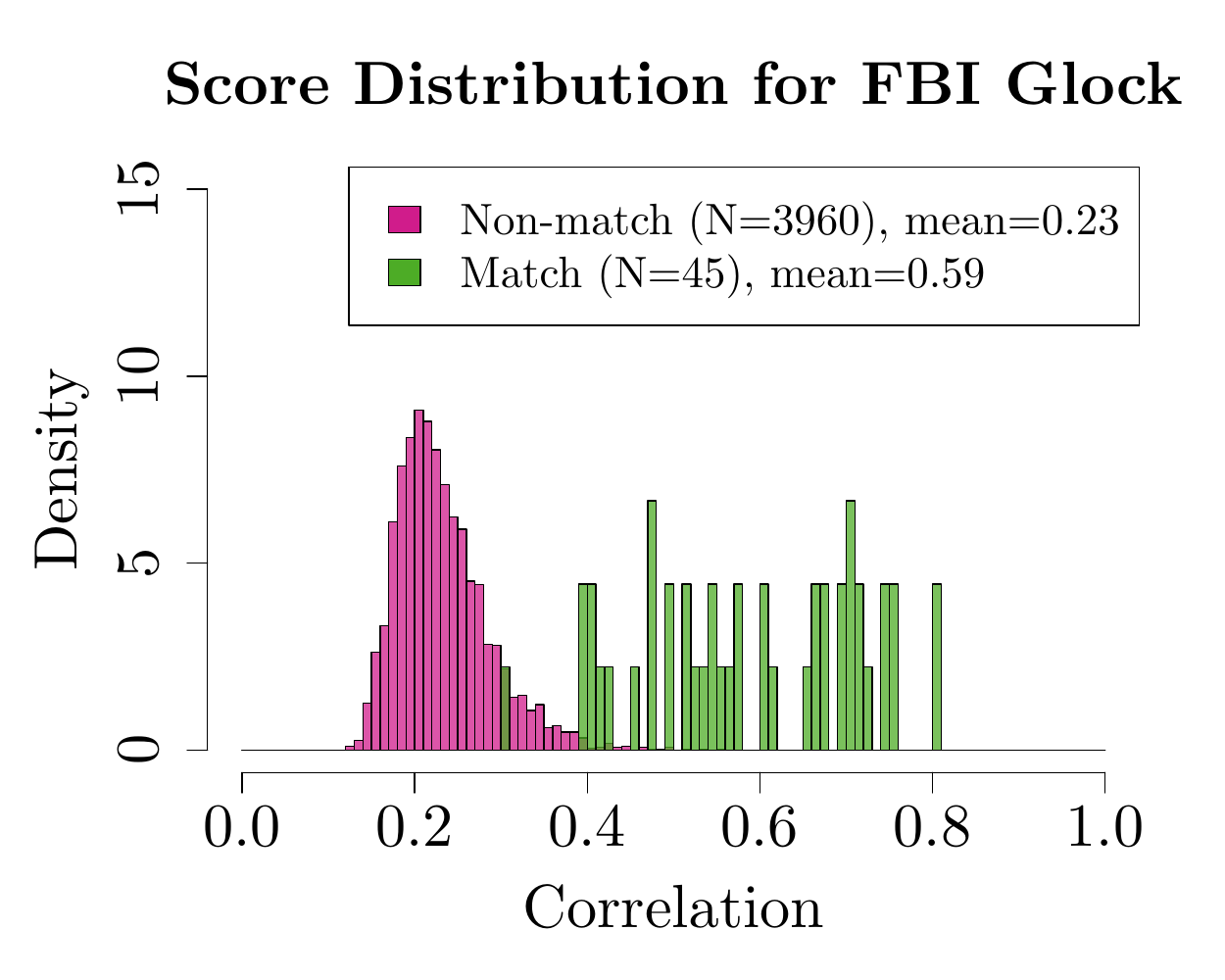}
		\caption{\label{fig:glock3D} FBI Glock}
	\end{subfigure}
	~
	\begin{subfigure}[t]{0.3\textwidth}
		\centering
		\includegraphics[width=\columnwidth]{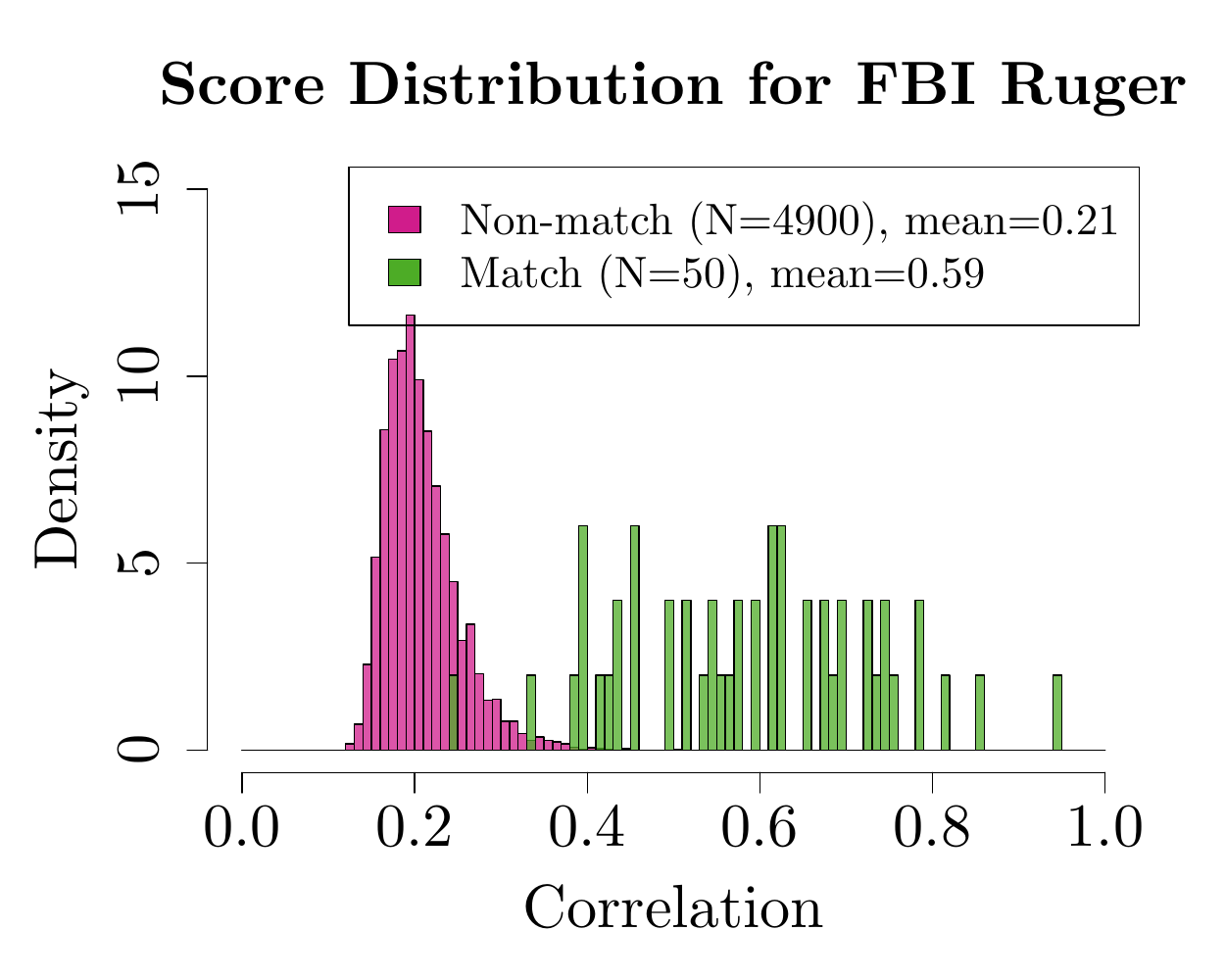}
		\caption{\label{fig:ruger3D} FBI Ruger}
	\end{subfigure}
	~
	\begin{subfigure}[t]{0.3\textwidth}
		\centering
		\includegraphics[width=\columnwidth]{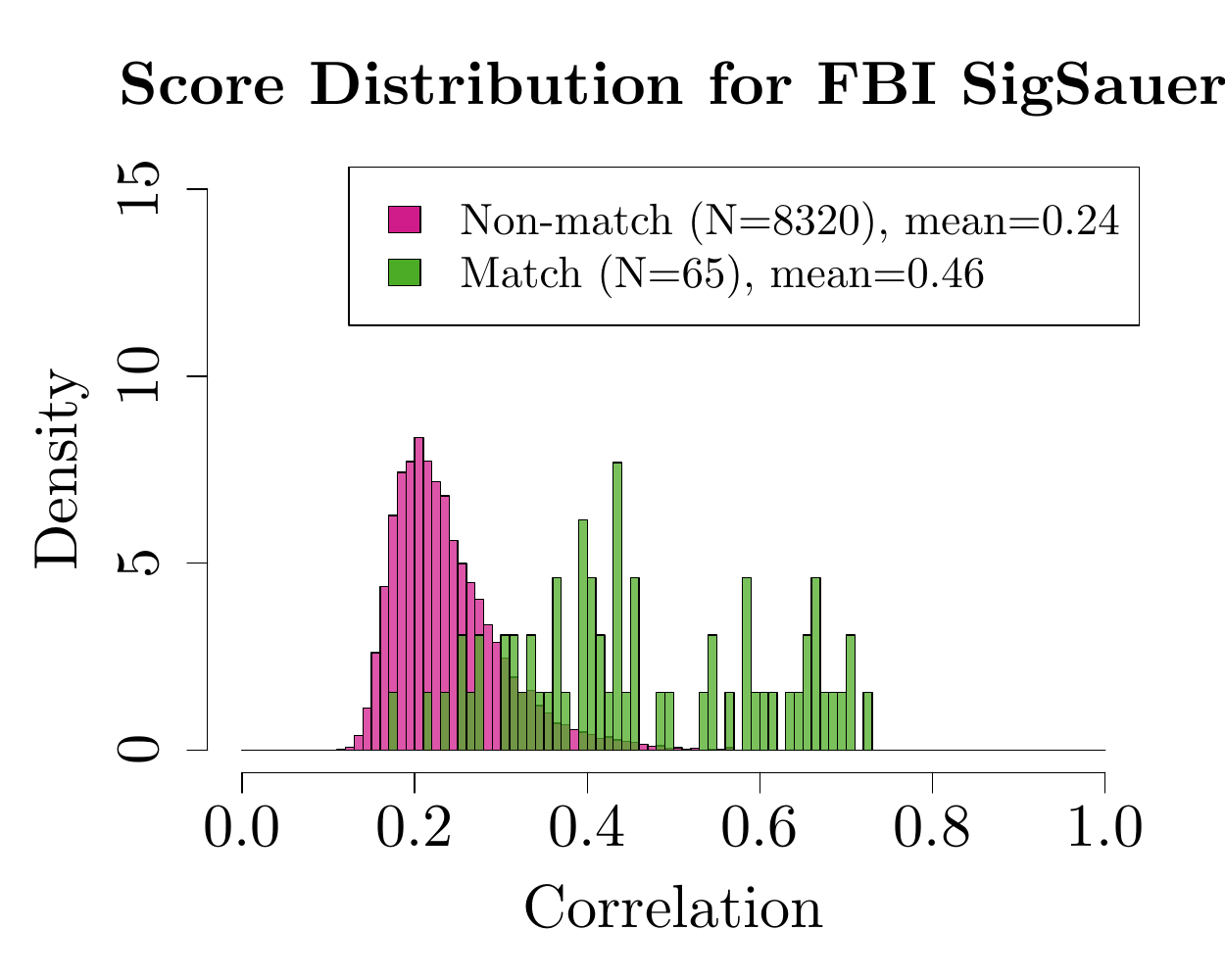}
		\caption{\label{fig:sig3D} FBI Sig Sauer}
	\end{subfigure}
	~
	\begin{subfigure}[t]{0.3\textwidth}
		\centering
		\includegraphics[width=\columnwidth]{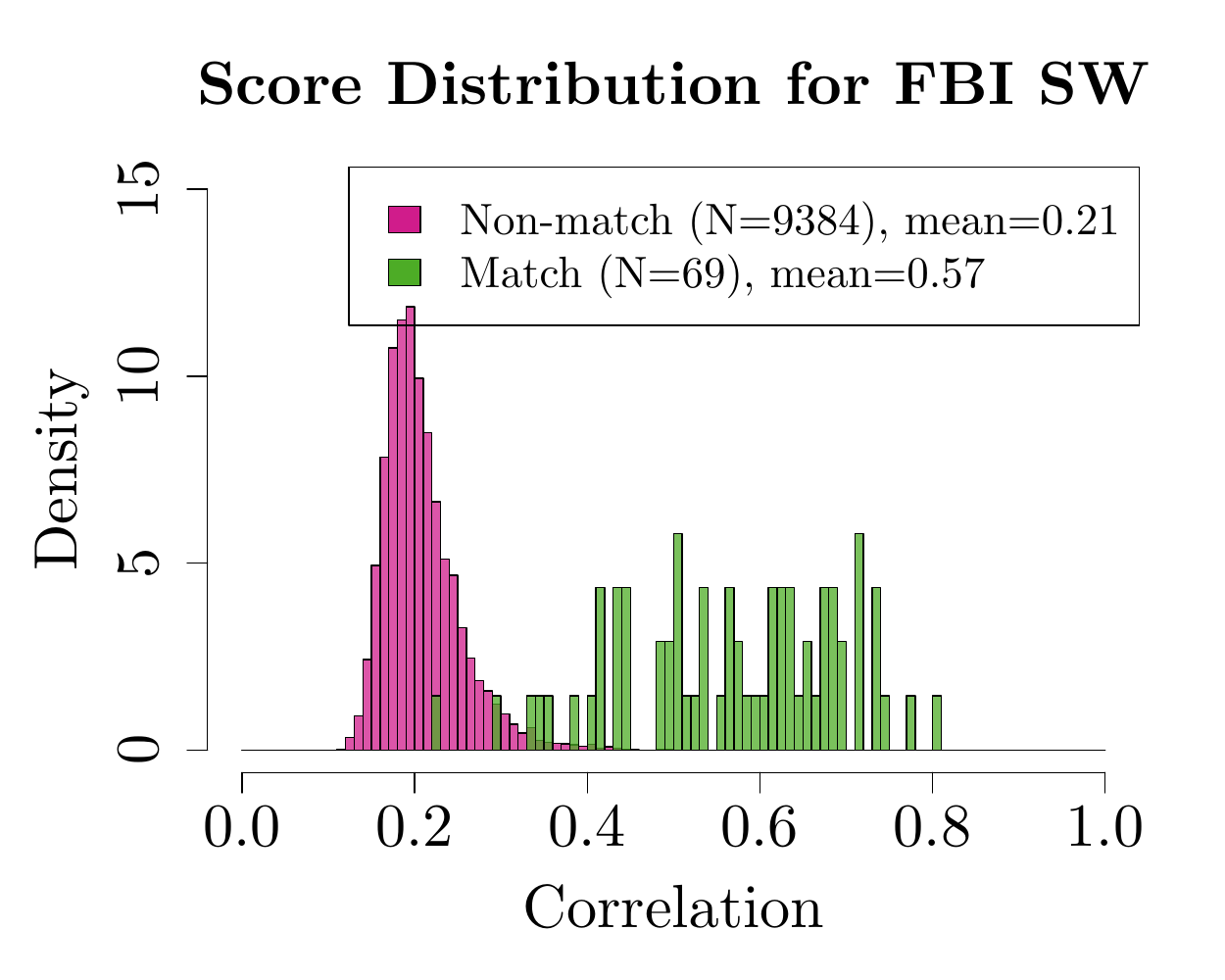}
		\caption{\label{fig:sw3D} FBI Smith \& Wesson}
	\end{subfigure}
~
	\centering
\begin{subfigure}[t]{0.3\textwidth}
	\centering
	\includegraphics[width=\columnwidth]{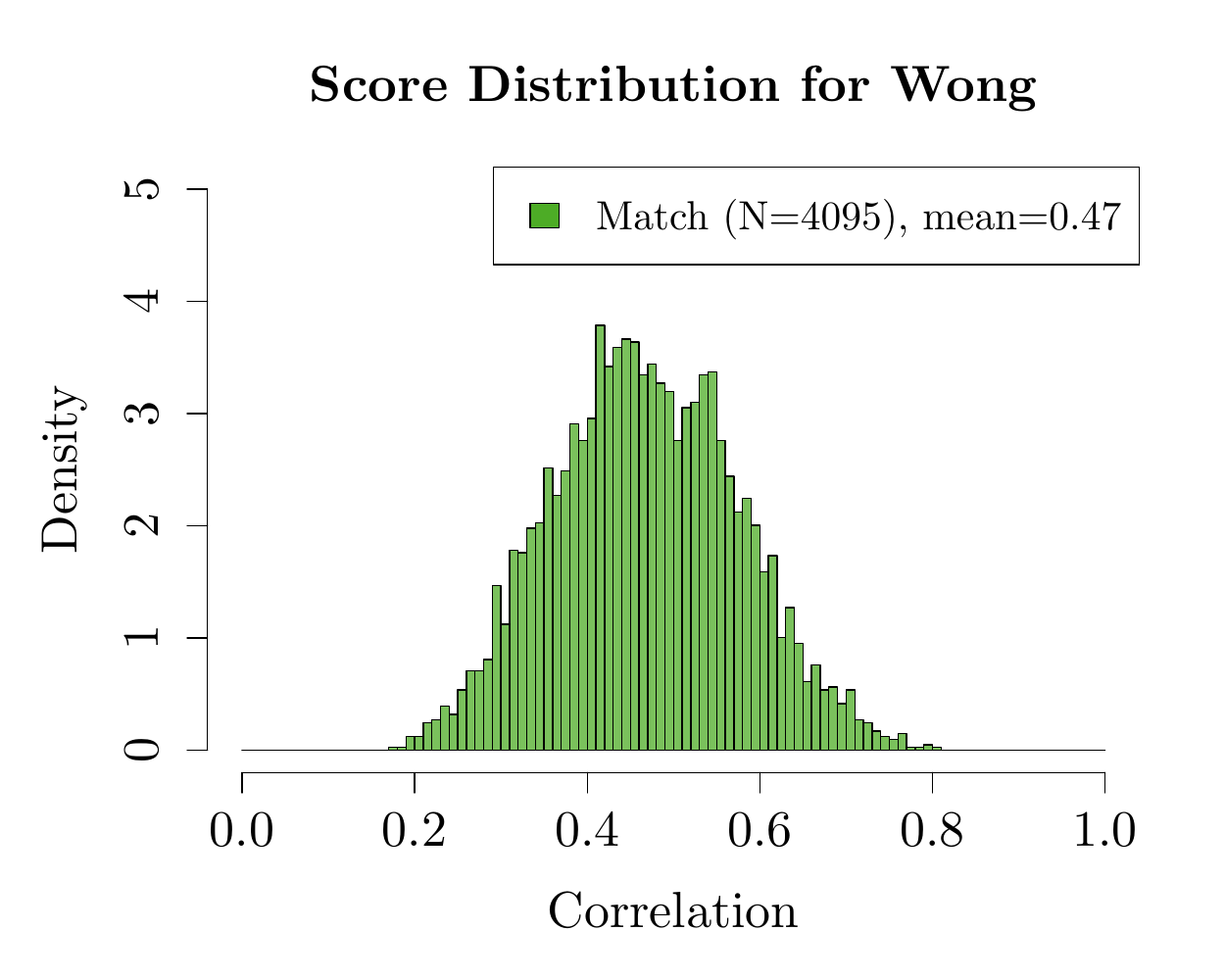} 
	\caption{\label{fig:wong3D} Cary Wong}
\end{subfigure}
	\hspace{10.5cm}
	
\end{figure}

\begin{figure}[!ht]
	\caption{\label{fig:pr3DbyStudy}
		Precision-recall plots by data set using 3D topographies. Area under the curve is reported in parentheses.}
	
	\centering
	\begin{subfigure}[t]{0.3\textwidth}
		\centering
		\includegraphics[width=\columnwidth]{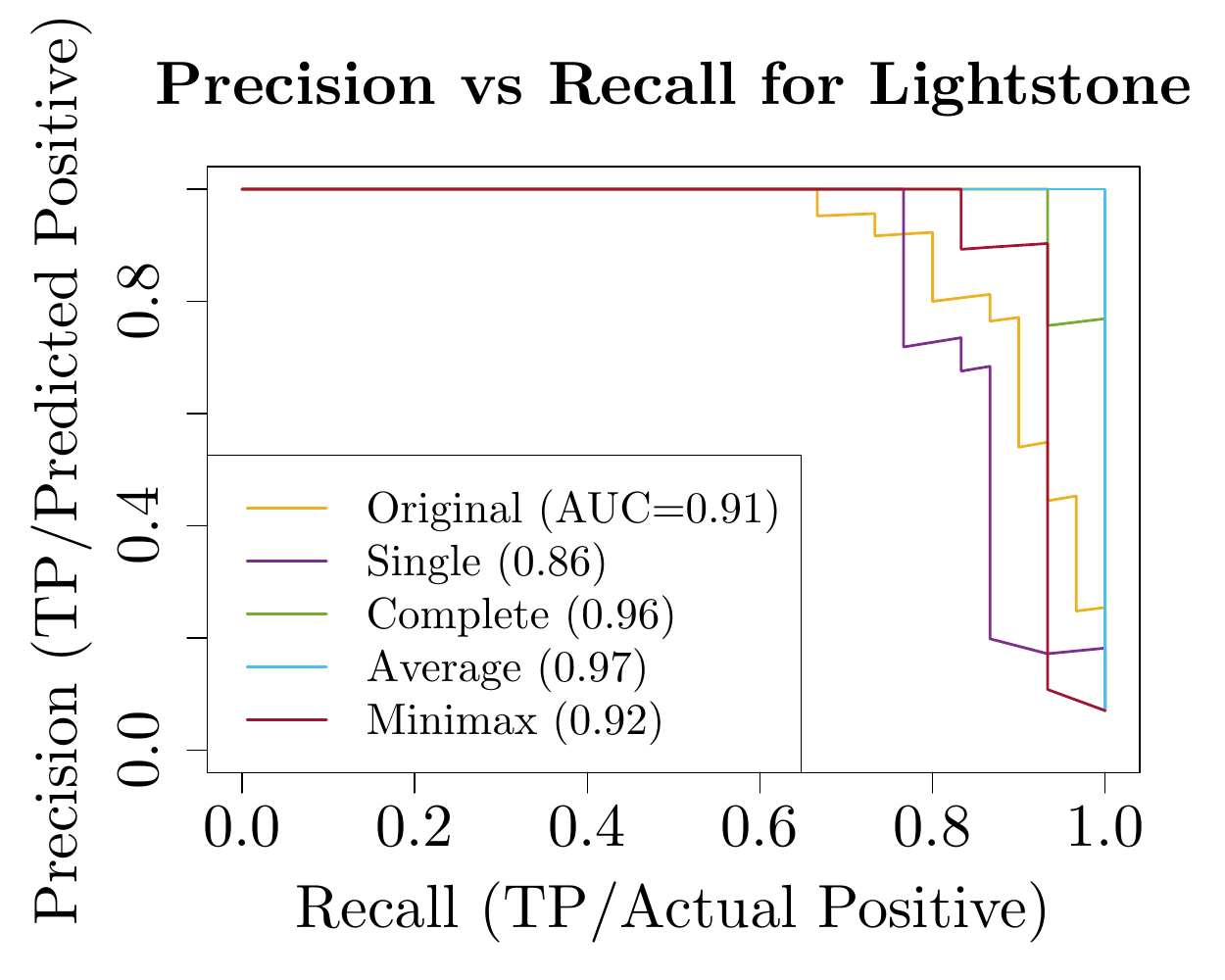}
		\caption{\label{fig:lightstone3Dpr} Lightstone (consecutively manufactured)}
	\end{subfigure}
	~
	\begin{subfigure}[t]{0.3\textwidth}
		\centering
		\includegraphics[width=\columnwidth]{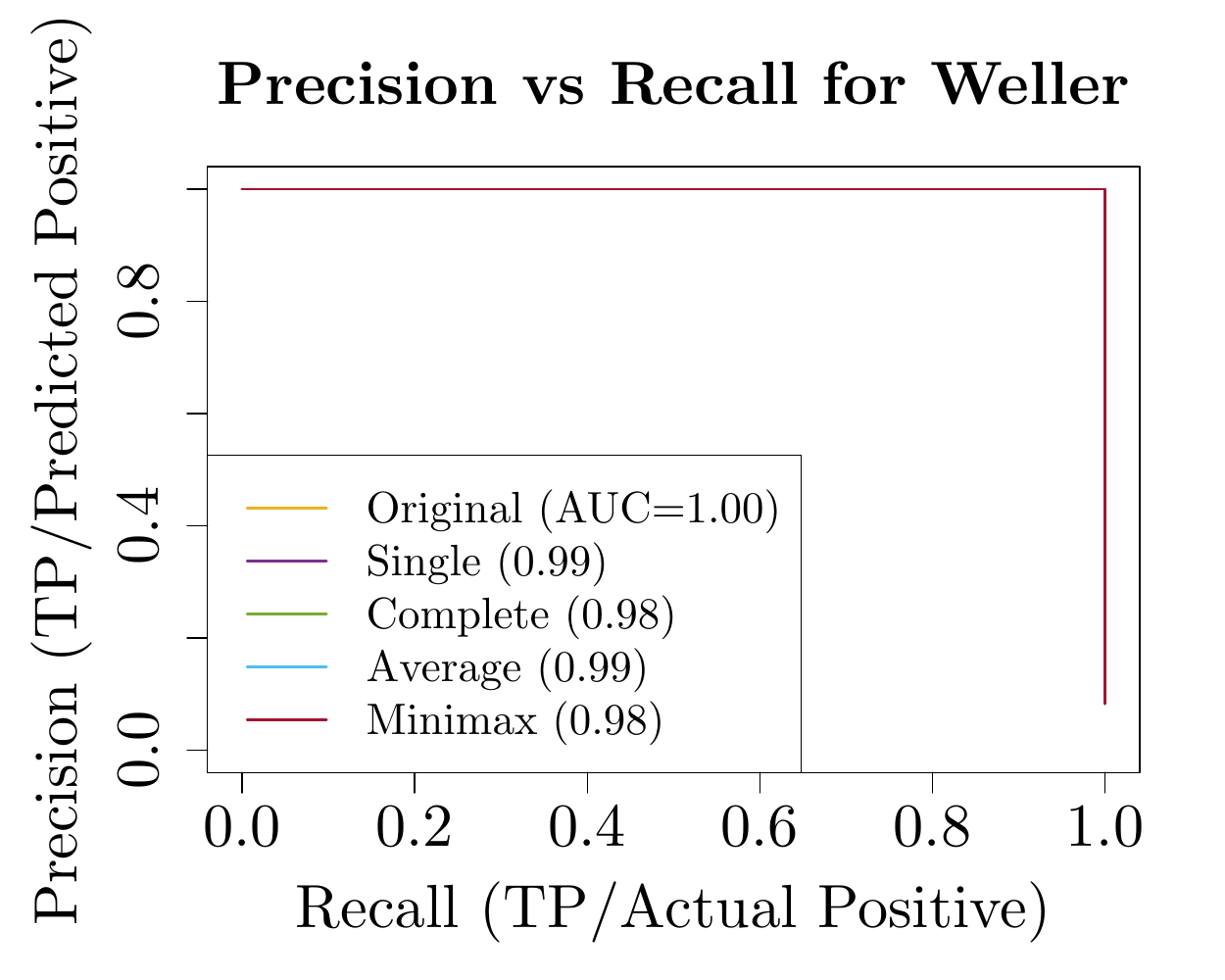}
		\caption{\label{fig:weller3Dpr} Weller (consecutively manufactured)}
	\end{subfigure}
	~
	\begin{subfigure}[t]{0.3\textwidth}
		\centering
		\includegraphics[width=\columnwidth]{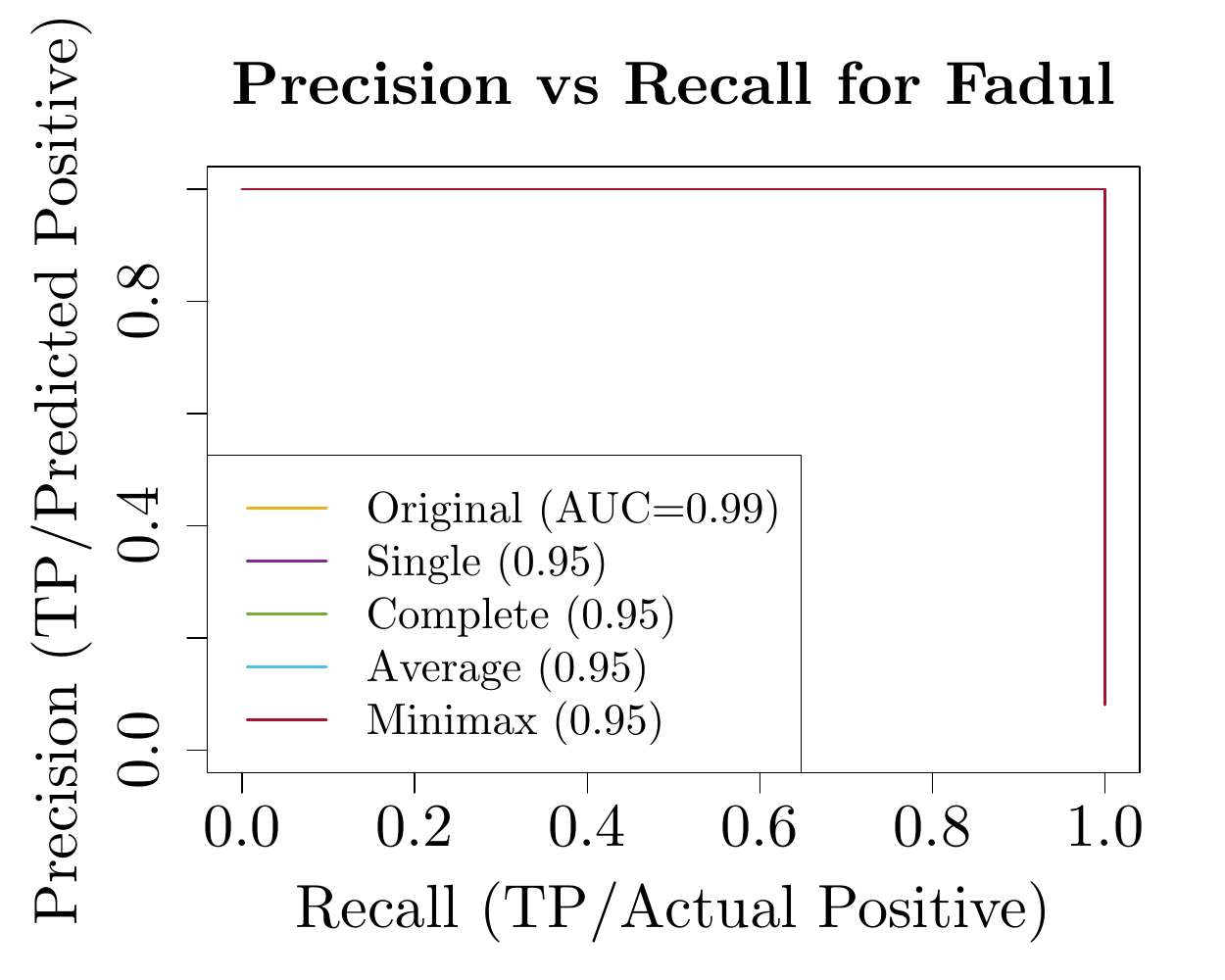}
		\caption{\label{fig:fadul3Dpr} Fadul (consecutively manufactured)}
	\end{subfigure}
	~
	\begin{subfigure}[t]{0.3\textwidth}
		\centering
		\includegraphics[width=\columnwidth]{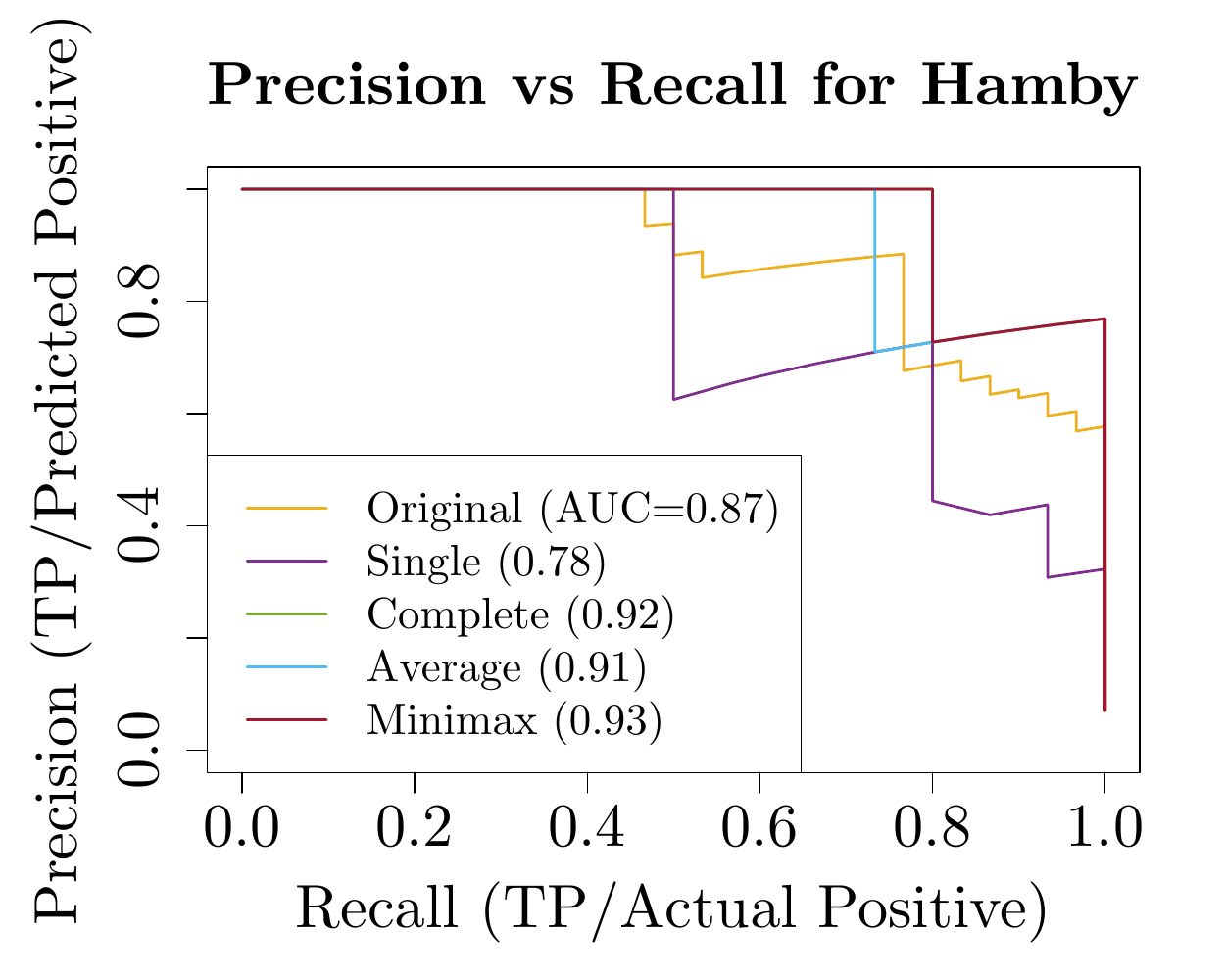}
		\caption{\label{fig:hamby3Dpr} Hamby (consecutively manufactured)}
	\end{subfigure}
	~
	\begin{subfigure}[t]{0.3\textwidth}
		\centering
		\includegraphics[width=\columnwidth]{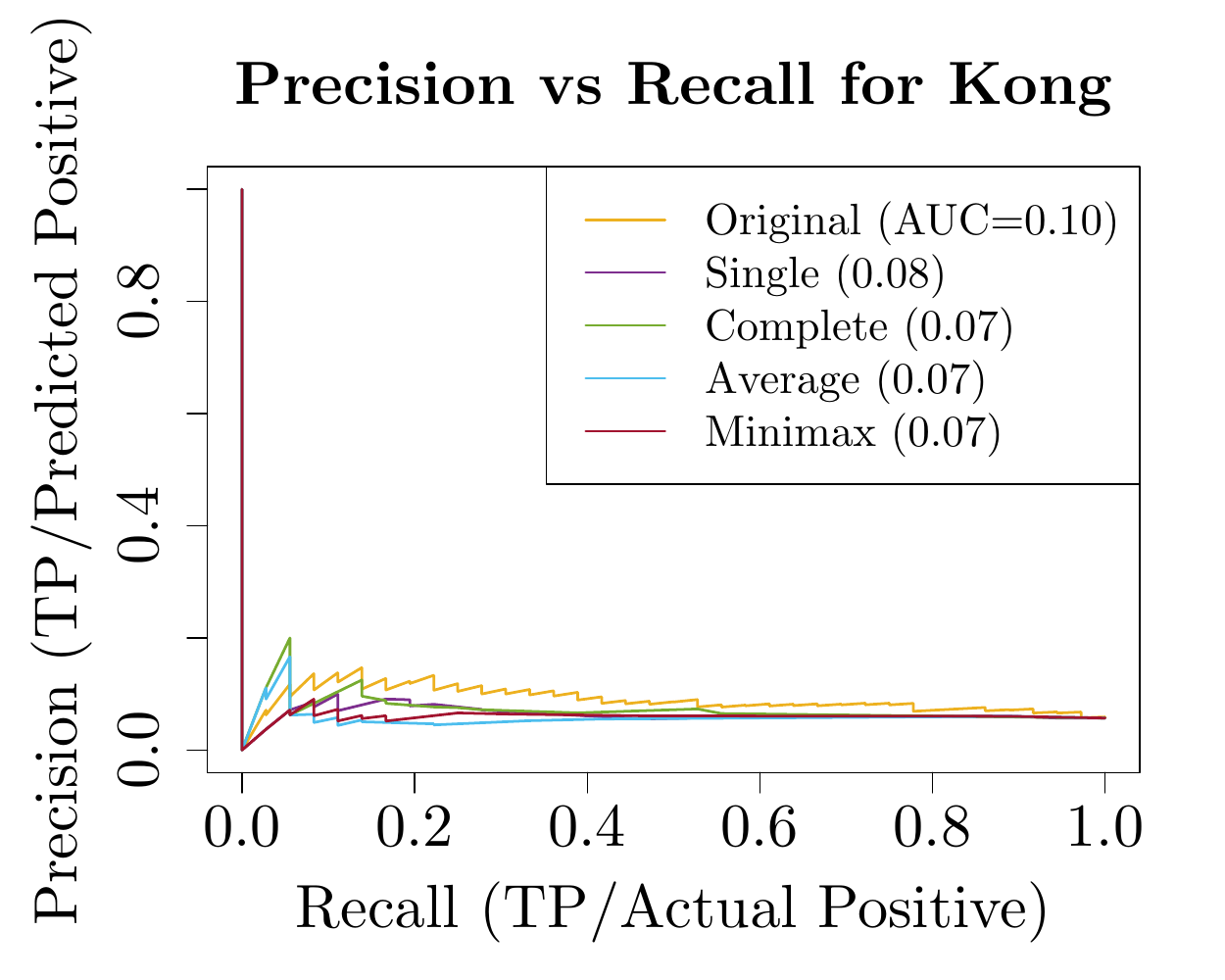}
		\caption{\label{fig:kong3Dpr} Kong}
	\end{subfigure}
	~
	\begin{subfigure}[t]{0.3\textwidth}
		\centering
		\includegraphics[width=\columnwidth]{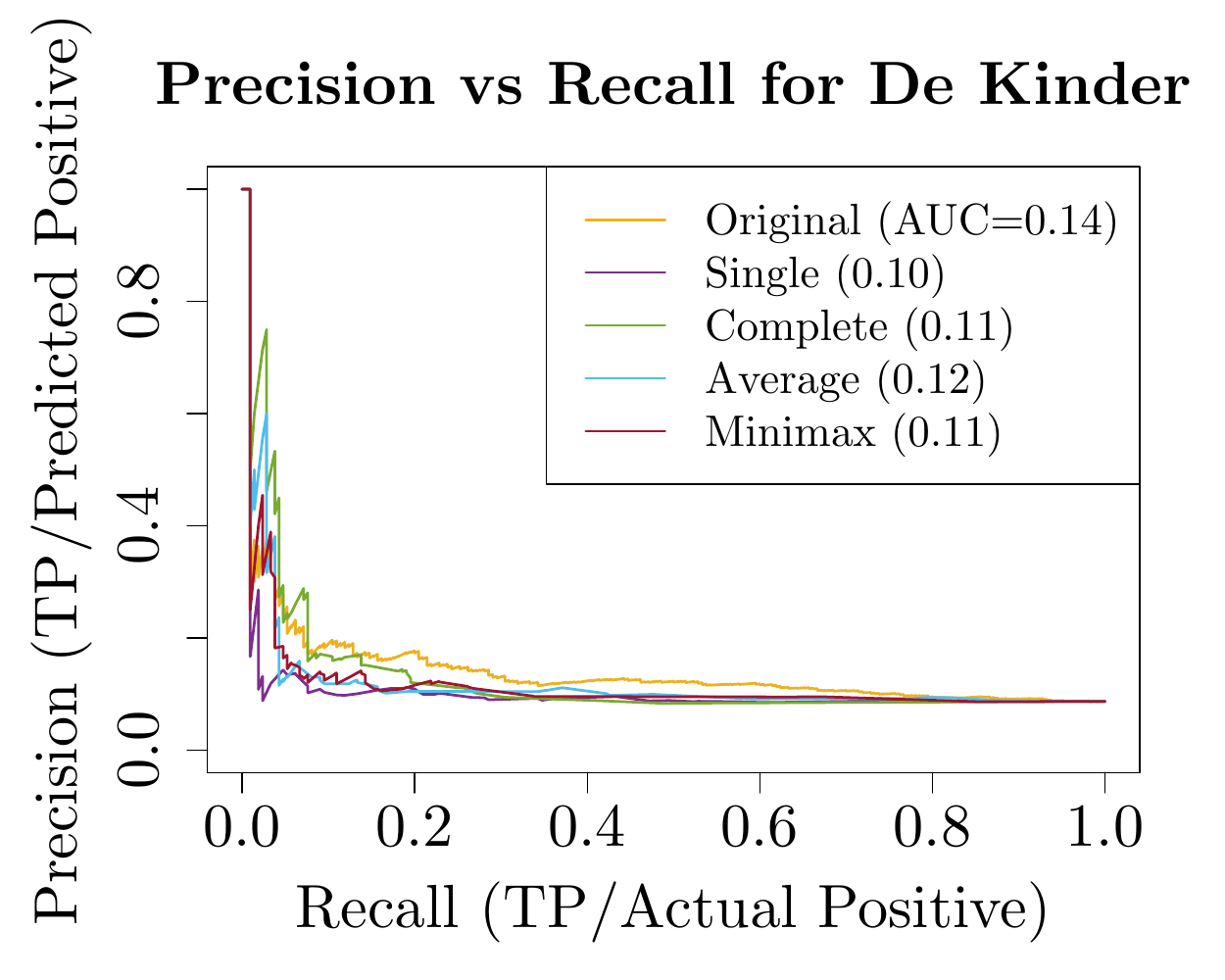}
		\caption{\label{fig:dekinder3Dpr} De Kinder}
	\end{subfigure}
	~
	\begin{subfigure}[t]{0.3\textwidth}
		\centering
		\includegraphics[width=\columnwidth]{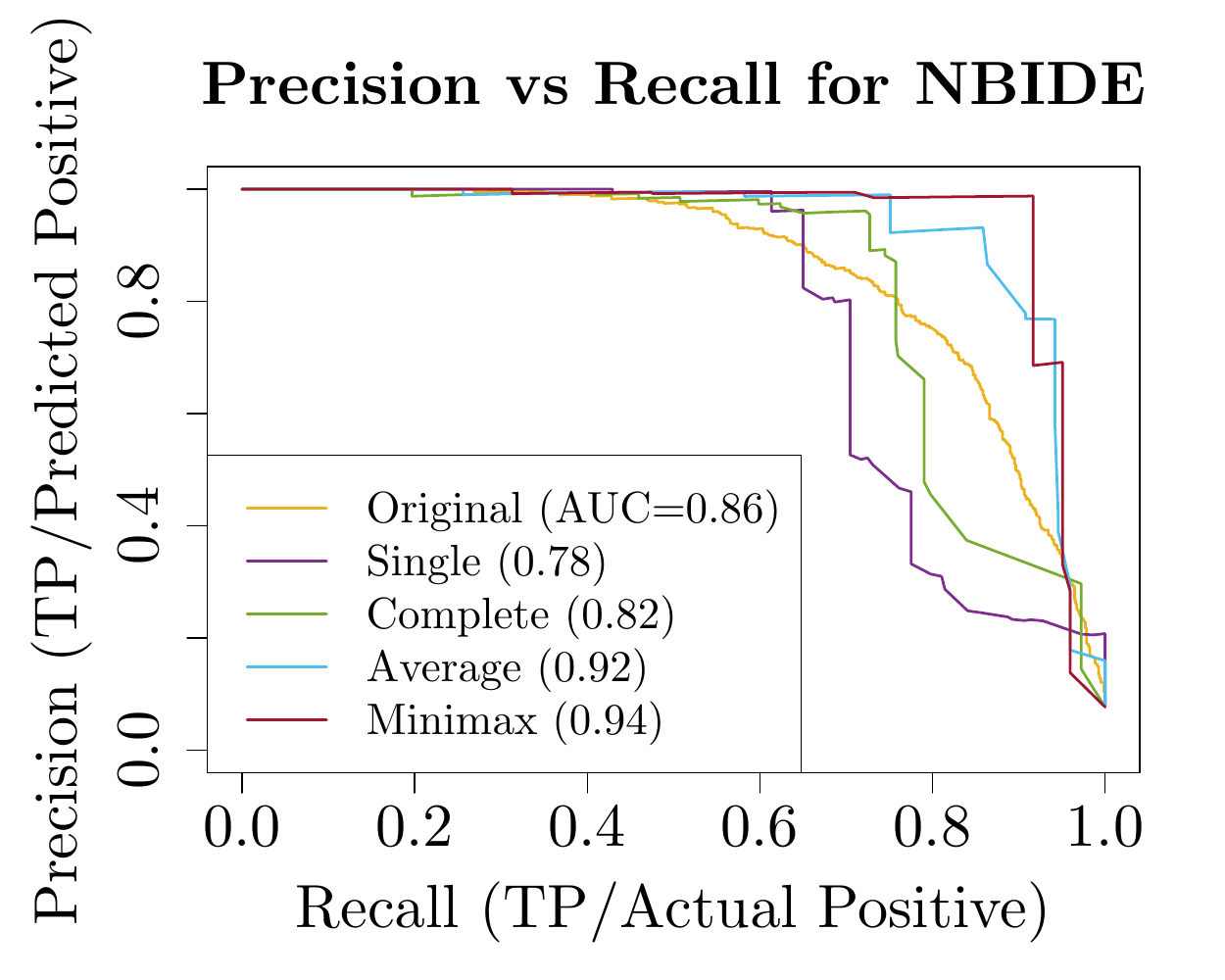}
		\caption{\label{fig:NBIDE3Dpr} NBIDE}
	\end{subfigure}
	~
	\begin{subfigure}[t]{0.3\textwidth}
		\centering
		\includegraphics[width=\columnwidth]{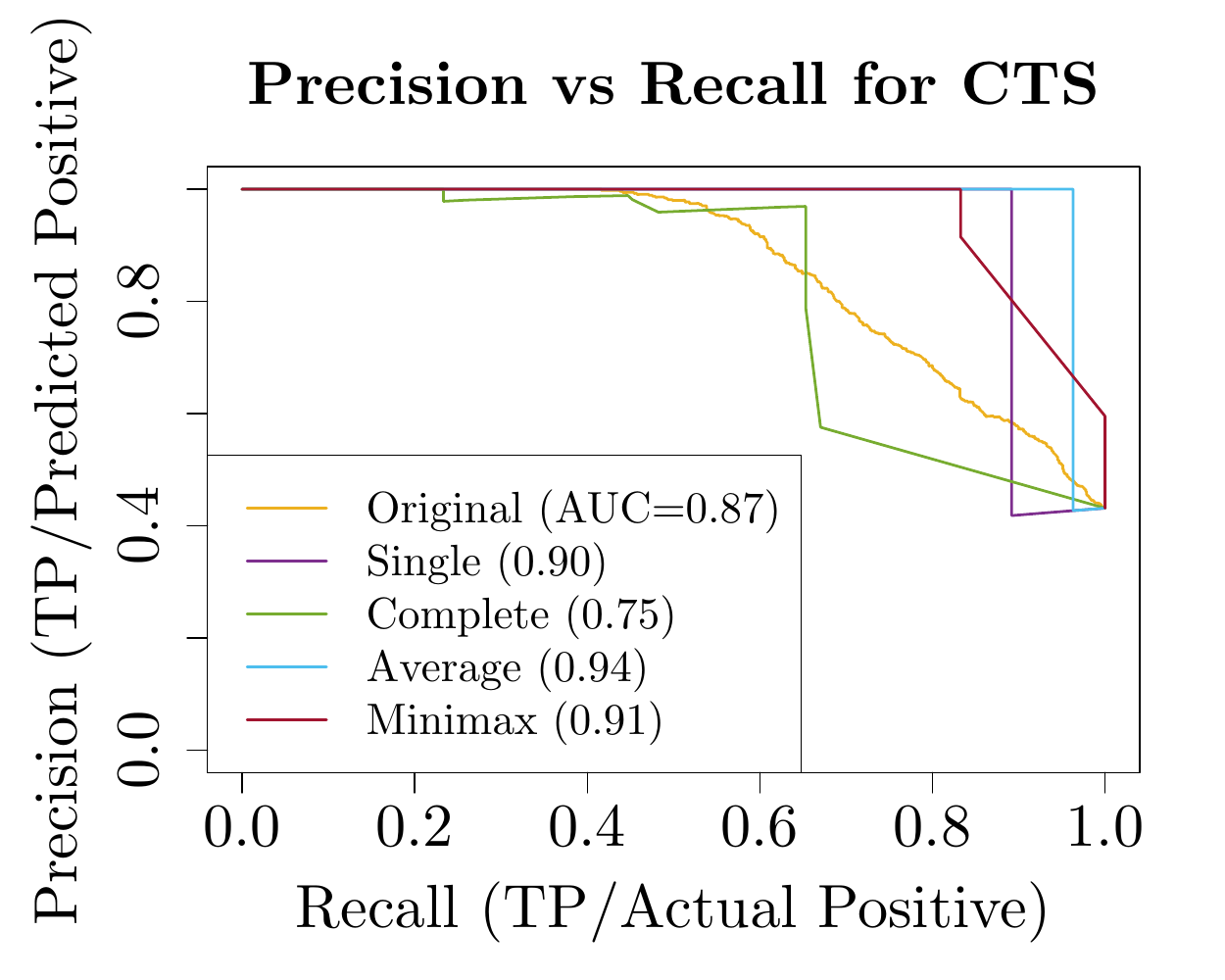}
		\caption{\label{fig:CTS3Dpr} CTS}
	\end{subfigure}
	~
	\begin{subfigure}[t]{0.3\textwidth}
		\centering
		\includegraphics[width=\columnwidth]{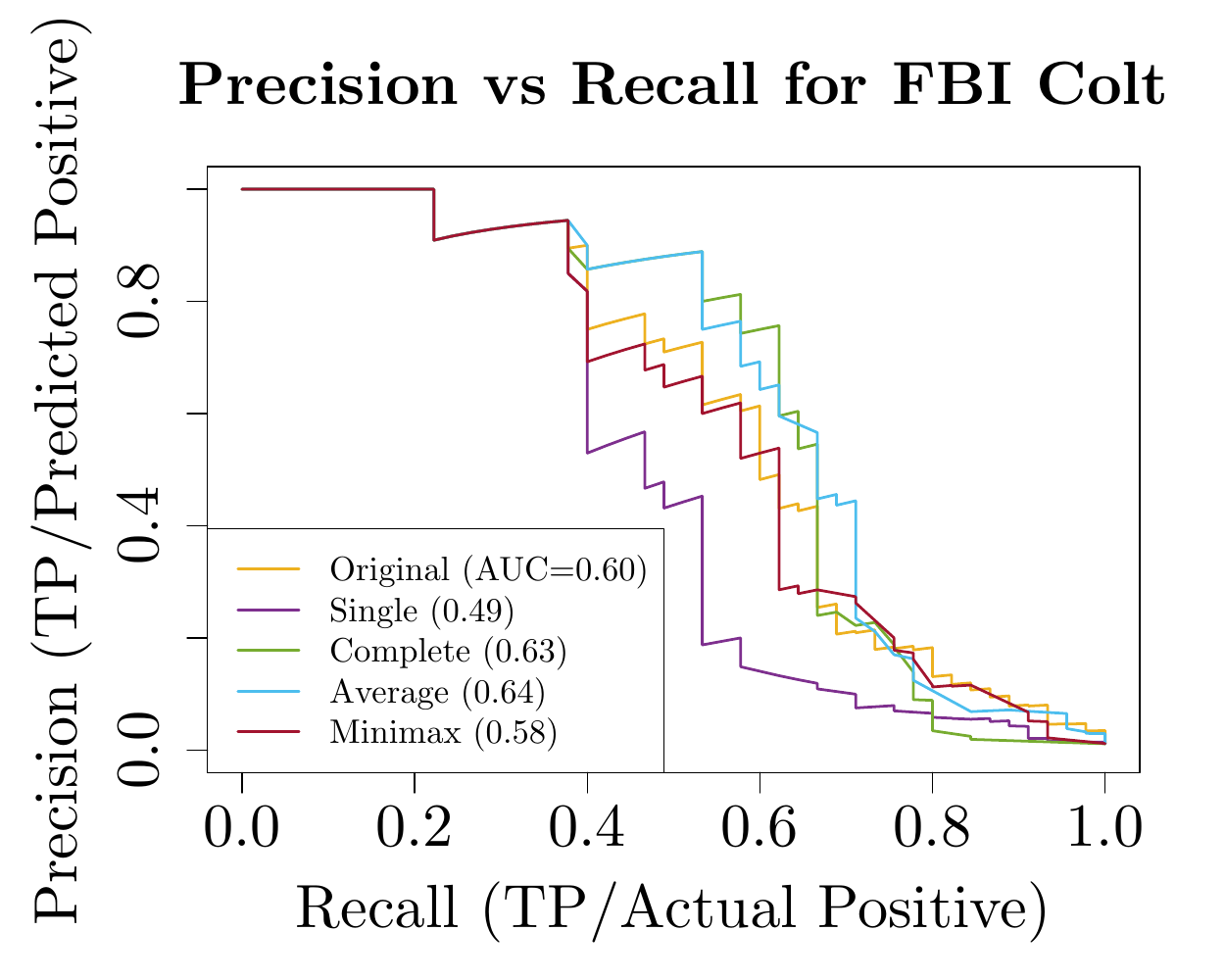}
		\caption{\label{fig:colt3Dpr} FBI Colt}
	\end{subfigure}
	~
	\begin{subfigure}[t]{0.3\textwidth}
		\centering
		\includegraphics[width=\columnwidth]{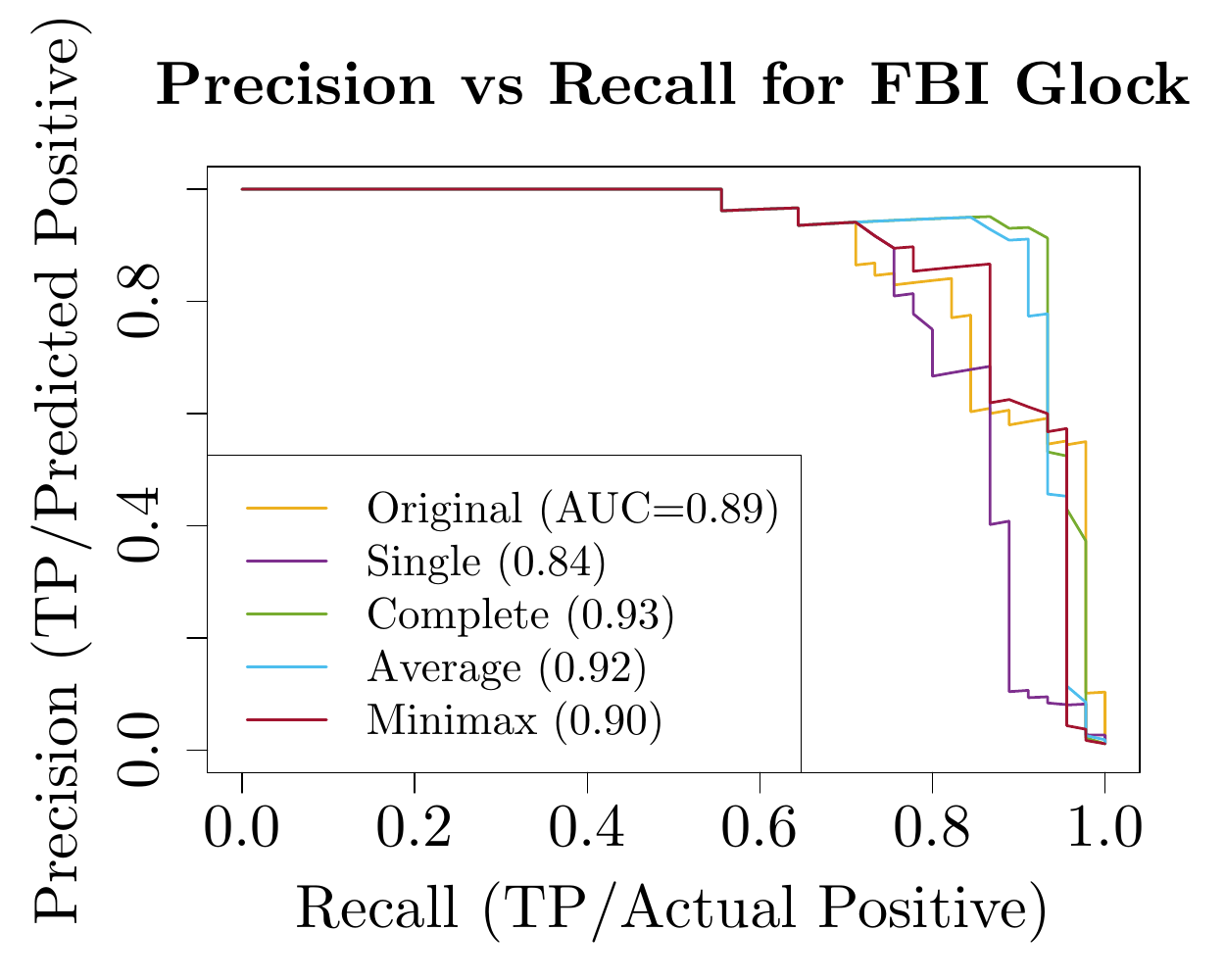}
		\caption{\label{fig:glock3Dpr} FBI Glock}
	\end{subfigure}
	~
	\begin{subfigure}[t]{0.3\textwidth}
		\centering
		\includegraphics[width=\columnwidth]{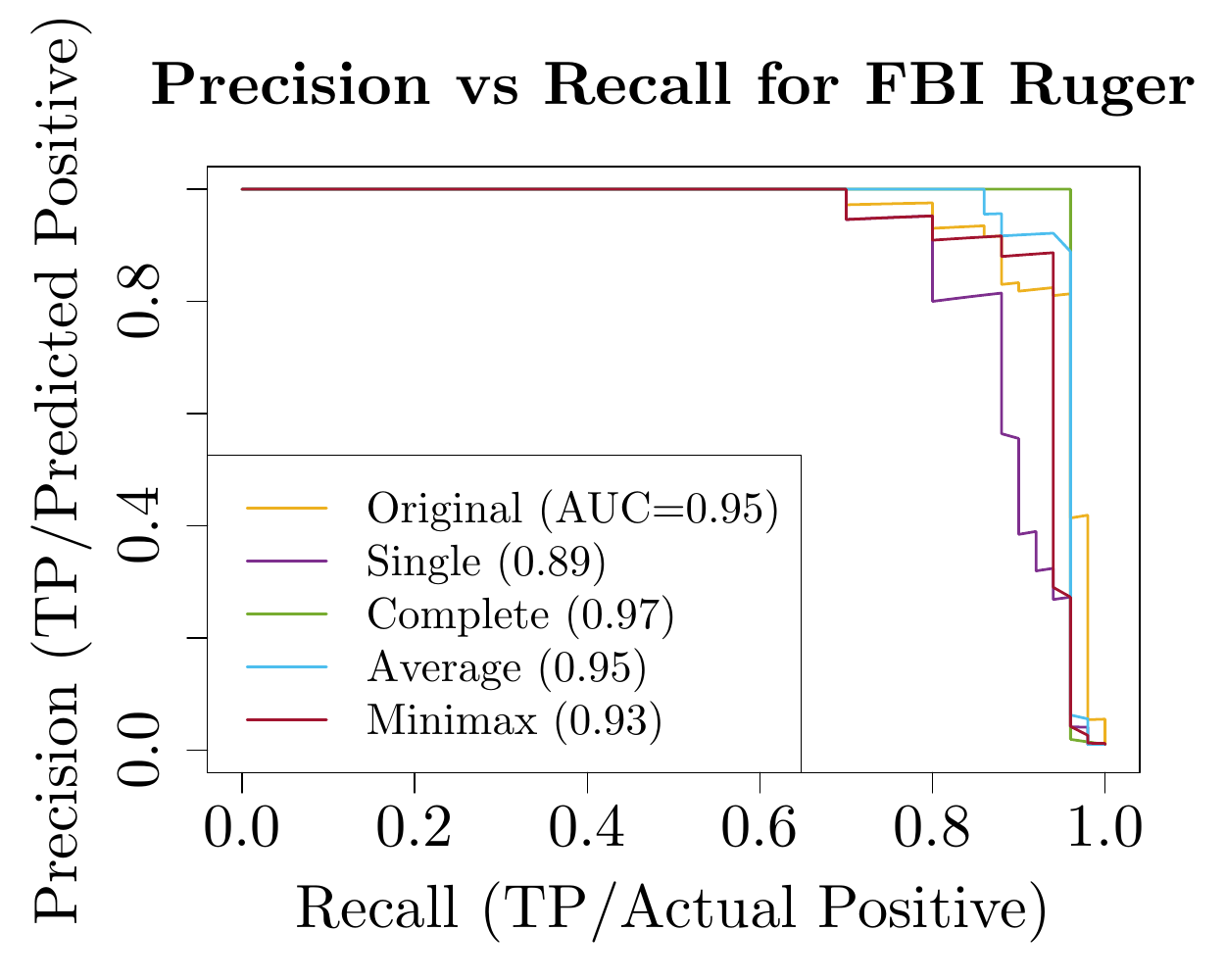}
		\caption{\label{fig:ruger3Dpr} FBI Ruger}
	\end{subfigure}
	~
	\begin{subfigure}[t]{0.3\textwidth}
		\centering
		\includegraphics[width=\columnwidth]{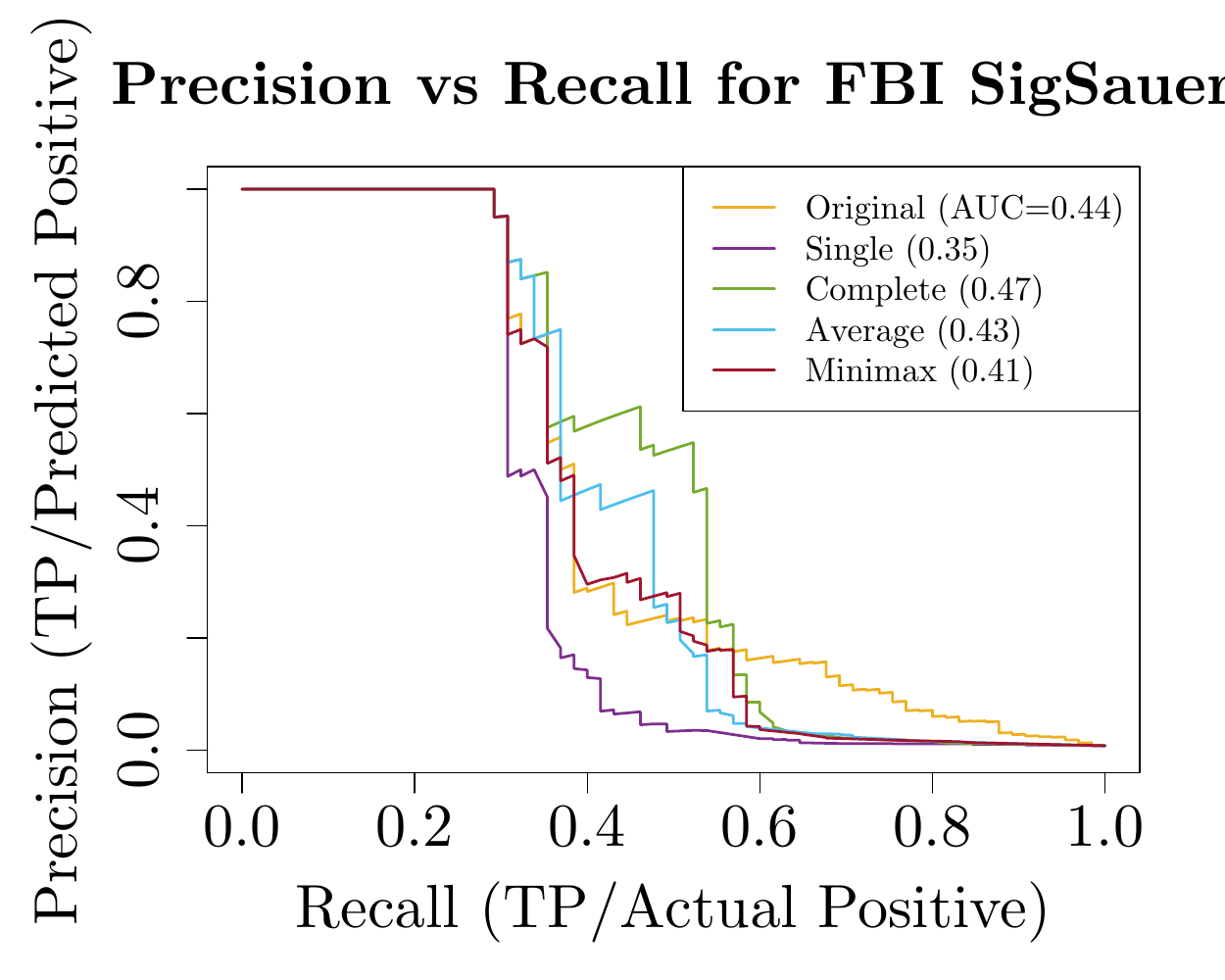}
		\caption{\label{fig:sig3Dpr} FBI Sig Sauer}
	\end{subfigure}
	~
	\begin{subfigure}[t]{0.3\textwidth}
		\centering
		\includegraphics[width=\columnwidth]{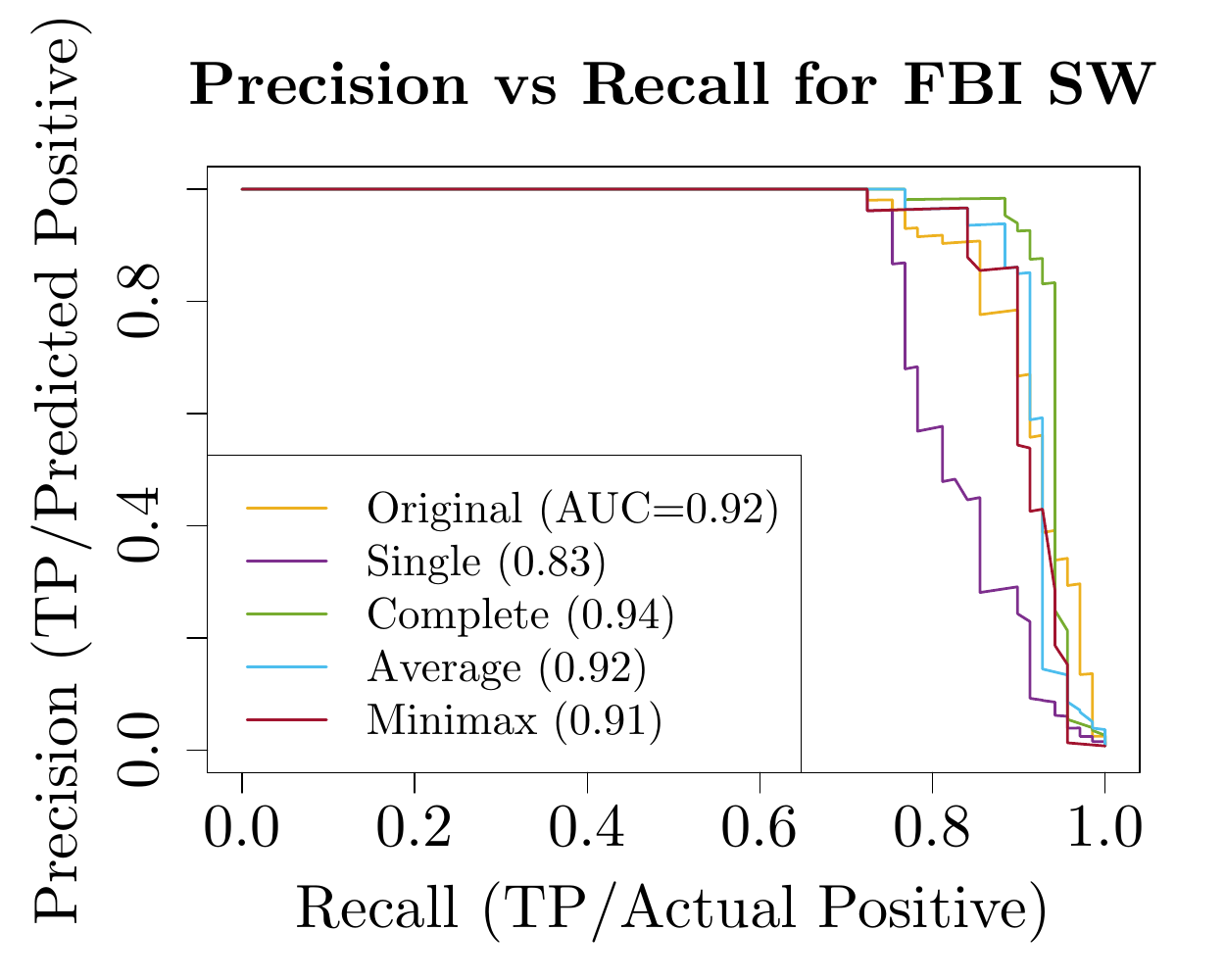}
		\caption{\label{fig:sw3Dpr} FBI Smith \& Wesson}
	\end{subfigure}
	\hspace{10.5cm}
	
\end{figure}

For each data set described in Table \ref{tab:firearmsData}, we run the complete pipeline as described in Section \ref{sec:method}. We perform all pairwise comparisons for each data set, evaluating the results both after the classification step and after the hierarchical clustering step. We first present distributions of similarity scores $\hat{s}$ for true matches and non-matches by data set in Figure \ref{fig:3DbyStudy}. We then look at precision-recall graphs in Figure \ref{fig:pr3DbyStudy}, with lines representing results both before and after hierarchical clustering. 


The results are organized by type of data set. Of the 14 data sets studied, four involve consecutively manufactured pistol slides firing the same ammunition. The pistol slide is a component of the firearm that includes the breech block, and is hence responsible for the breechface marks. It had been suggested that pistol slides that were produced successively might have similar patterns and imperfections that persist through the manufacturing process \citep[see e.g.,][]{Lightstone2010}, so cartridge cases obtained using such slides might be more difficult to differentiate. \cite{Lightstone2010} found that firearms examiners did not in fact have trouble comparing such cartridge cases, because even though the molds that were used to produce the slides did have markings that carried over through the manufacturing process, the individual characteristics on each slide were sufficient for examiners to make correct identifications to the individual slide. Here one can come to the same conclusion: in Figures \ref{fig:3DbyStudy}\subref{fig:lightstone3D} to \ref{fig:3DbyStudy}\subref{fig:hamby3D} the method produces higher similarity scores for most comparisons belonging to the same pistol slide, than those involving different slides. All the plots show a very good separation between the true match and non-match distributions. At least for the combinations of firearm and ammunition studied, having pistol slides that are consecutively manufactured does not appear to pose a problem for automatic identification. 

Now consider the Kong and De Kinder data sets, which involve multiple copies of the same firearm. Kong uses 12 Smith \& Wesson 10-10's, firing Fiocchi ammunition. One might expect results to mimic those that involve consecutively manufactured slides, but this does not turn out to be the case; results are far worse. De Kinder on the other hand, uses 10 Sig Sauer P226's, firing six different brands of ammunition, and the results also show extremely poor separation between match and non-match scores. From the Kong study, manually looking at examples of non-matched pairs producing very high scores (over .6), one observation is that the markings appear as small scratches, and are very much less pronounced compared to those in Figure \ref{fig:mechanism}\subref{fig:marks}, say. It has been noted by firearms examiners that some guns mark more poorly than others, leading to difficult comparisons, and this could be an example of this phenomenon. 

Next, results for the NBIDE, CTS and FBI data sets are in Figures \ref{fig:3DbyStudy}\subref{fig:NBIDE3D} to \ref{fig:3DbyStudy}\subref{fig:sw3D}. These all involve either multiple ammunition or firearm types. The five FBI data sets use various Colts, Glocks, Rugers, Sig Sauers and Smith \& Wessons, and only Remington ammunition, while CTS uses Rugers and Smith \& Wesson firearms and Federal ammunition. NBIDE uses different firearms with Remington, Winchester, Speer and PMC ammunition. For all these data sets, there is a range of separation between the match and non-match scores. Gun brands that have the best performance look to be Glock, Ruger and Smith \& Wesson. 

Finally, the Cary Wong data set involves a single type of firearm (Ruger P89) and ammunition (Winchester), but the gun is fired 2000 times. Cartridge cases 25-1000 at intervals of 25 test fires are imaged, as well as the first 10 and the 1001st test fire, giving a total of 91 images. The purpose of this study was to investigate if marks persisted over multiple fires, say, over the lifespan of a gun that is heavily used. The results in 
show a range of $\hat{s}$ values. This suggests that we may be able to identify some but not all cartridge cases to this particular firearm, in its entire lifespan. 

%
%
%


Precision-recall graphs before and after clustering are in Figure \ref{fig:pr3DbyStudy}. Note that the Cary Wong data set is excluded from this presentation because there are no non-matched pairs in this data set. The results in Figure \ref{fig:pr3DbyStudy} corroborate the visual findings presented in Figure \ref{fig:3DbyStudy}, and provide a convenient numerical summary of performance. We find that the clustering step results in better AUCs for data sets that are not at the extremes in terms of performance (either very poor or close to perfect): the larger differences are for Hamby (.87 to .93), NBIDE (.86 to .94), CTS (.87 to .94), FBI Colt (.6 to .64). Overall, very good performance is achieved in 9 of the 13 data sets, where AUCs are over .9 using at least one linkage method. One might hypothesize as to why clustering improves performances. Minimax linkage, for example, selects a cluster center for each cluster, and members of the cluster need to be similar to this center but not necessarily to other members of the cluster. It is plausible that the cluster center is a particularly well-marked cartridge case, while other cluster members might have areas that are less well-marked. 

We have so far been agnostic about the selection of a similarity cutoff to designate predicted matches and non-matches. This might be of interest in some situations. For example, one might be interested in selecting pairs above a cutoff for further manual investigation. Alternatively, a cutoff could be used to make a conclusion in criminal cases on whether a person should be implicated in a crime. The choice of cutoff depends on the goal and desired tradeoff between false negatives and false positives. In the first investigative case a lower cutoff might be set to ensure high recall, while in the second case a much higher cutoff might be necessary, since the costs associated with falsely implicating a suspect are disproportionately high.

\subsubsection{Final clusters}	\label{sssec:finalClusters}
Suppose that for comparison with examiner proficiency tests, we are interested in producing a disambiguated data set where images are clustered based on whether they share a common source. This requires the selection of a cutoff as well as a linkage method. Figure \ref{fig:pr3DbyStudy} shows the range of values of precision and recall that can be achieved using various cutoffs and different linkage methods. For the purposes of illustration we treat false negatives and false positives equally, and choose a cutoff that maximizes precision and recall; this corresponds to the value at the bend of the precision-recall curve. As for the linkage method, examining results from the different data sets, it appears that minimax and average linkage have good performance across the board. In addition, minimax linkage has the advantage of being interpretable, so minimax linkage is selected to generate final clusters. Using minimax linkage, a cutoff that works well for all data sets is around .4, so we select this value.

We illustrate the results using the NBIDE data set; one can repeat this analysis for any data set. 
The NBIDE data set consists of a total of 144 images from 12 different guns. If the algorithm worked perfectly, one would expect to see 12 clusters of size 12. Instead, the cluster sizes are shown in Table \ref{tab:NBIDEclustSizes}. Now, comparing to ground truth, all 6 clusters of 12 are correct, and correspond to 4 Ruger firearms, and 2 Sig Sauers. The remaining 6 firearms comprise of combining two or more of the smaller clusters, with one exception: one cluster of size 4 mistakenly puts a cartridge case from a Smith \& Wesson firearm in a cluster containing cartridge cases from a different Smith \& Wesson firearm. In terms of performance by gun brand, it appears that from best to worst are Ruger, Sig Sauer, and then Smith \& Wesson.
		
\begin{center}
	\captionof{table}{\label{tab:NBIDEclustSizes}Cluster sizes for the NBIDE data set in 3D, after hierarchical clustering using minimax linkage with a cutoff of .4. Perfect results would be 12 clusters of size 12.} 
	\centering
	\begin{tabular}{lcccccccc}
		\hline 
		Cluster Size & 1 & 3 & 4 & 5  & 6 & 7 & 9&  12  \\
		\hline 
		Count   & 2 & 5  & 4      &  1    &   3 &  1 & 1 & 6 \\	
		\hline
	\end{tabular}
\end{center}


\subsection{Comparison between 2D and 3D} \label{ssec:2Dvs3D}
In Section \ref{sec:data} we explained the two types of technology currently available for producing digital representations of the data: 2D reflectance images, and 3D topographical measurements. We mentioned that NIST has advocated for the use of 3D topographies. In this section we investigate how the individual pairwise comparison results differ for the two types of images, to investigate if the 3D representation does indeed produce superior results. The results presented in Section \ref{ssec:3D} involve 14 different data sets. The same data are available in 2D, with the exception of CTS and Todd Weller. CTS data are completely unavailable in 2D, while a subset of 50 out of the 95 Weller images are available in 2D. All available data are used for comparing the two different types of data.

For processing 2D images, we use methodology from \cite{Tai2018}. Full details are in the paper; the main differences are that for 2D data we used edge detection and image processing operations to select the breechface marks, whereas in 3D, we make use of differences in depth measurements. Processing 2D data also involves the additional step of removing outliers. The resolution used for comparisons in 2D was 10$\mu m$, compared to 25$\mu m$ for 3D topographies.

\begin{figure}[!ht]
	\caption{\label{fig:2Dvs3D}
		2D scores vs 3D scores by data set.}
	
	\centering
	\begin{subfigure}[t]{0.23\textwidth}
		\centering
		\includegraphics[width=\columnwidth]{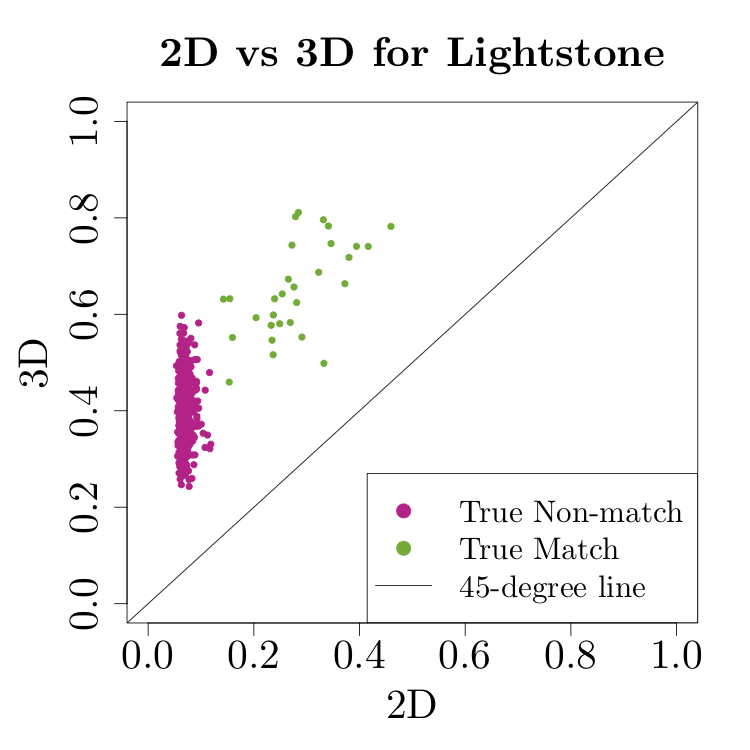}
		\caption{\label{fig:lightstone2Dvs3D} Lightstone (consecutively manufactured)}
	\end{subfigure}
	~
	\begin{subfigure}[t]{0.23\textwidth}
		\centering
		\includegraphics[width=\columnwidth]{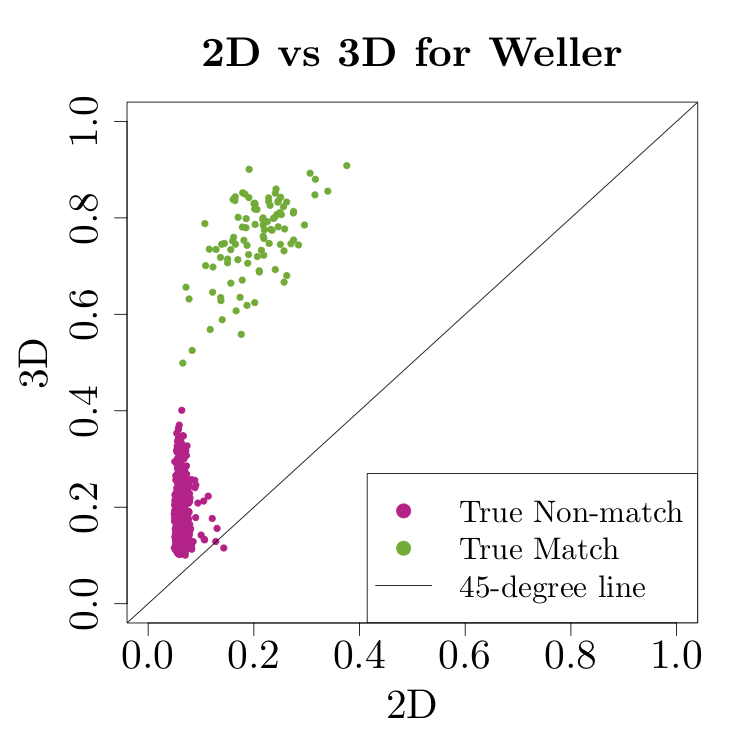}
		\caption{\label{fig:weller2Dvs3D} Weller (consecutively manufactured)}
	\end{subfigure}
	~
	\begin{subfigure}[t]{0.23\textwidth}
		\centering
		\includegraphics[width=\columnwidth]{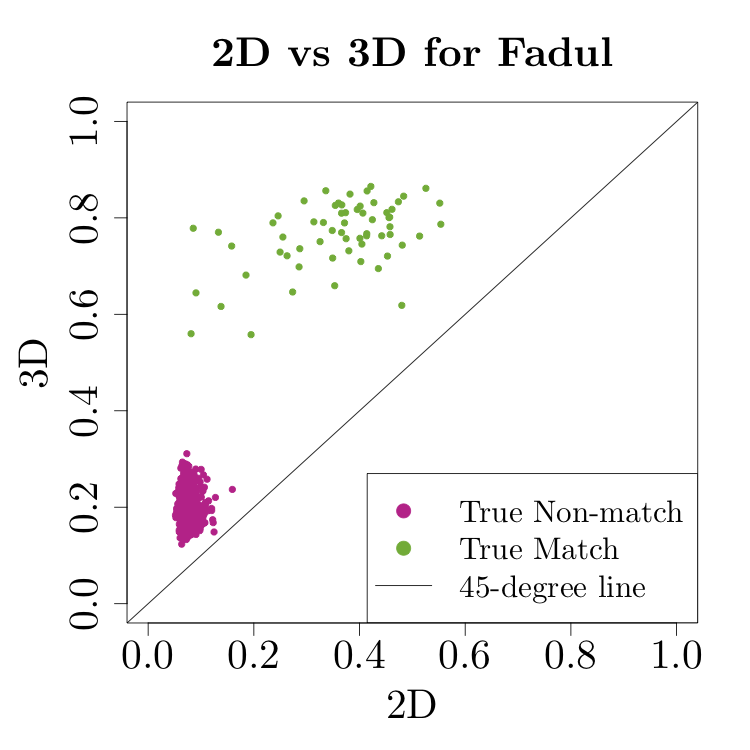}
		\caption{\label{fig:fadul2Dvs3D} Fadul (consecutively manufactured)}
	\end{subfigure}
	~
	\begin{subfigure}[t]{0.23\textwidth}
		\centering
		\includegraphics[width=\columnwidth]{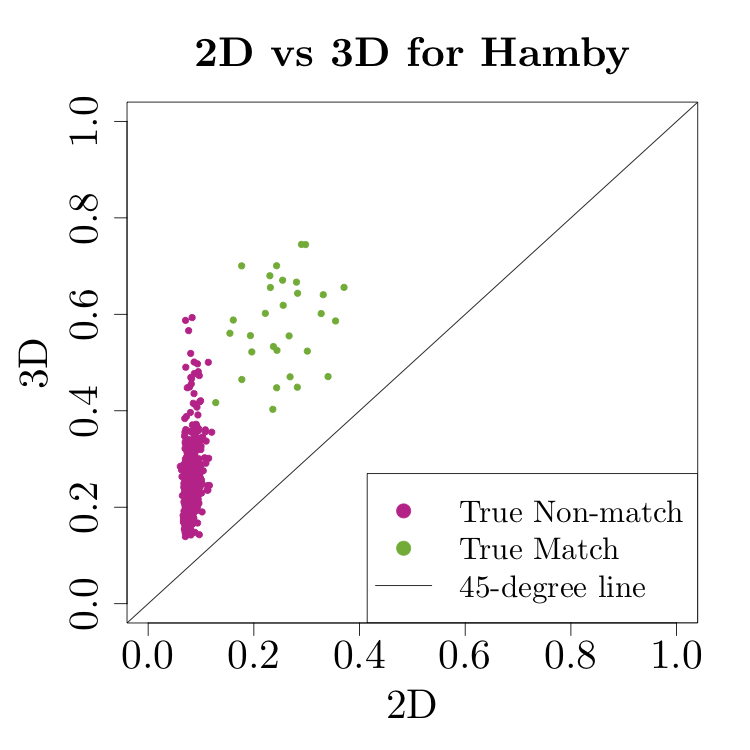}
		\caption{\label{fig:hamby2Dvs3D} Hamby (consecutively manufactured)}
	\end{subfigure}
	~
	\begin{subfigure}[t]{0.23\textwidth}
		\centering
		\includegraphics[width=\columnwidth]{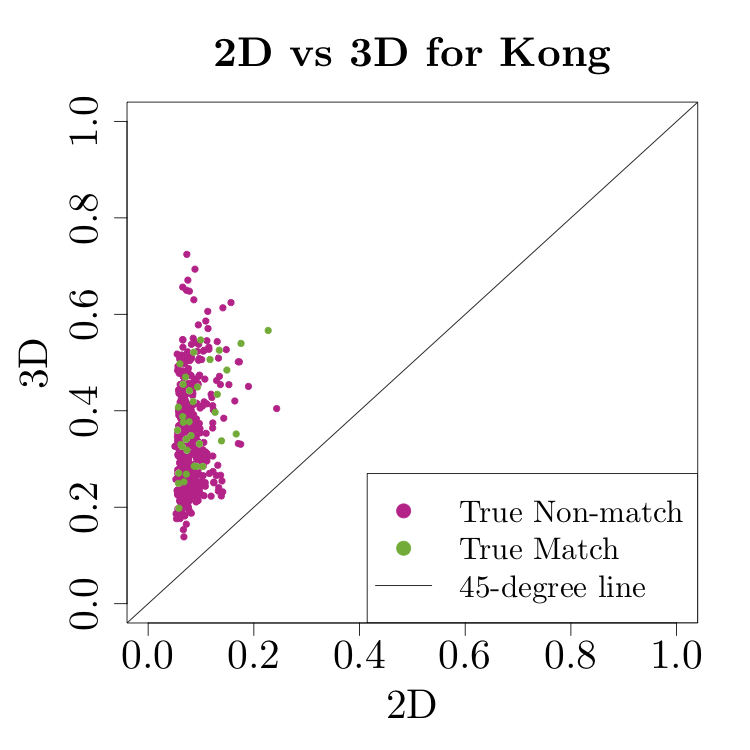}
		\caption{\label{fig:kong2Dvs3D} Kong}
	\end{subfigure}
	~
	\begin{subfigure}[t]{0.23\textwidth}
		\centering
		\includegraphics[width=\columnwidth]{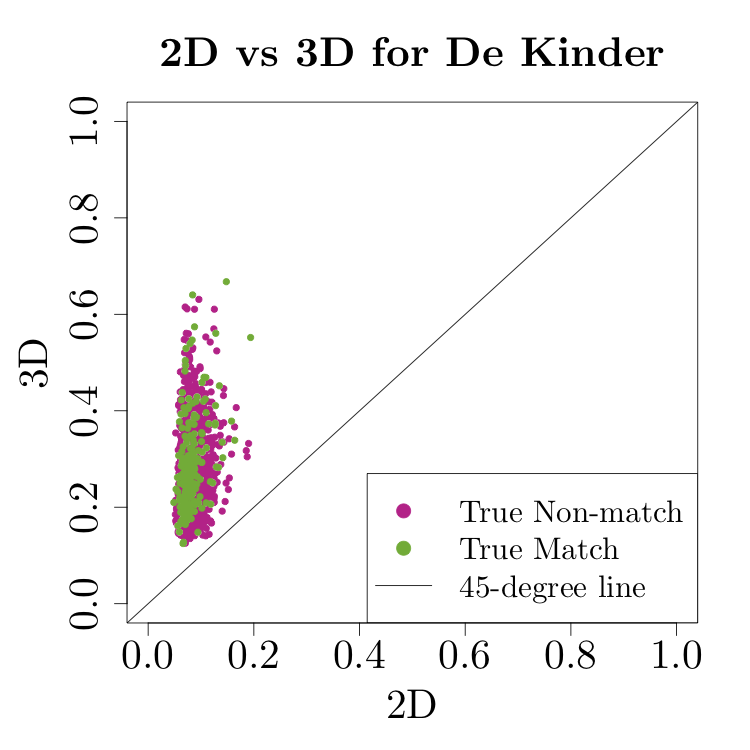}
		\caption{\label{fig:dekinder2Dvs3D} De Kinder}
	\end{subfigure}
	~
	\begin{subfigure}[t]{0.23\textwidth}
		\centering
		\includegraphics[width=\columnwidth]{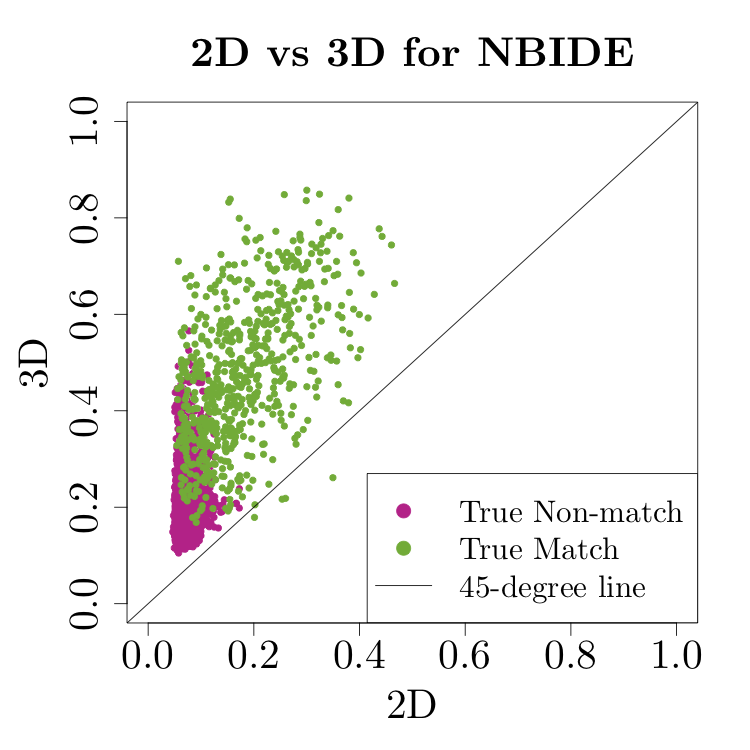}
		\caption{\label{fig:NBIDE2Dvs3D} NBIDE}
	\end{subfigure}
	~
	\begin{subfigure}[t]{0.23\textwidth}
		\centering
		\includegraphics[width=\columnwidth]{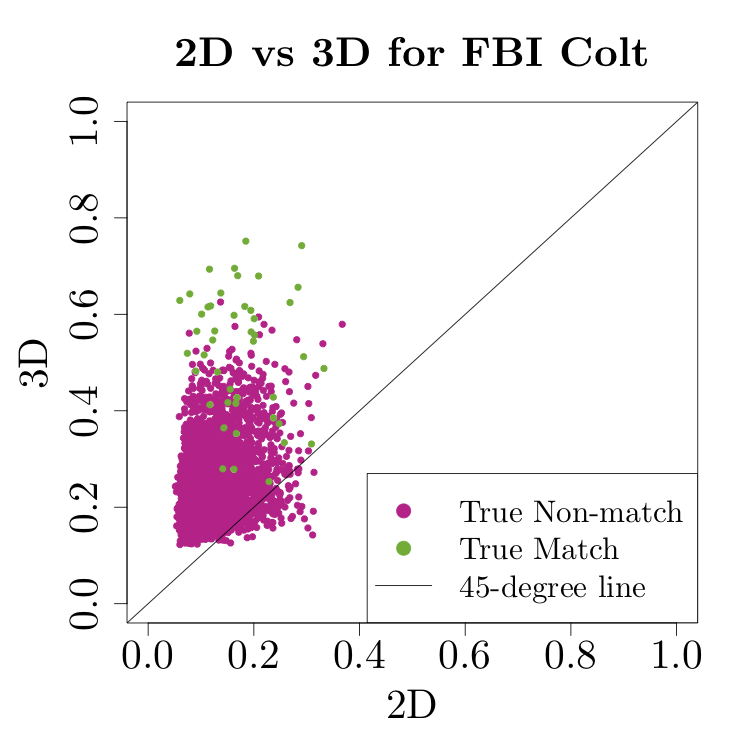}
		\caption{\label{fig:colt2Dvs3D} FBI Colt}
	\end{subfigure}
	~
	\begin{subfigure}[t]{0.23\textwidth}
		\centering
		\includegraphics[width=\columnwidth]{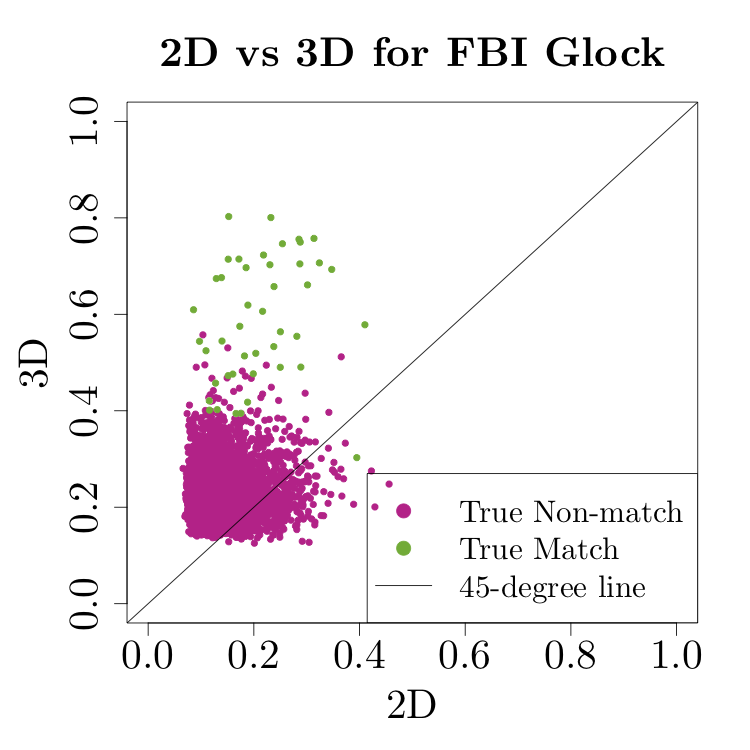}
		\caption{\label{fig:glock2Dvs3D} FBI Glock}
	\end{subfigure}
	~
	\begin{subfigure}[t]{0.23\textwidth}
		\centering
		\includegraphics[width=\columnwidth]{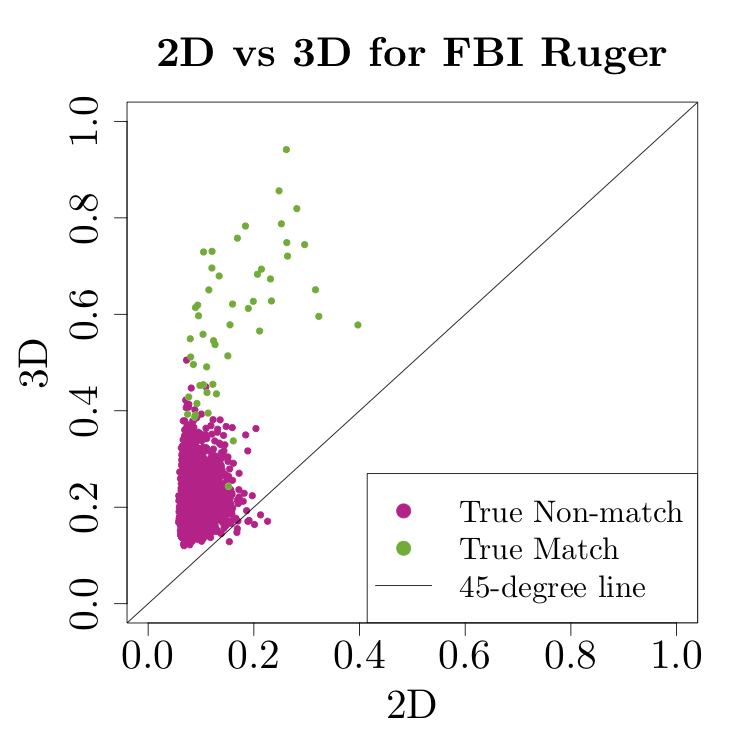}
		\caption{\label{fig:ruger2Dvs3D} FBI Ruger}
	\end{subfigure}
	~
	\begin{subfigure}[t]{0.23\textwidth}
		\centering
		\includegraphics[width=\columnwidth]{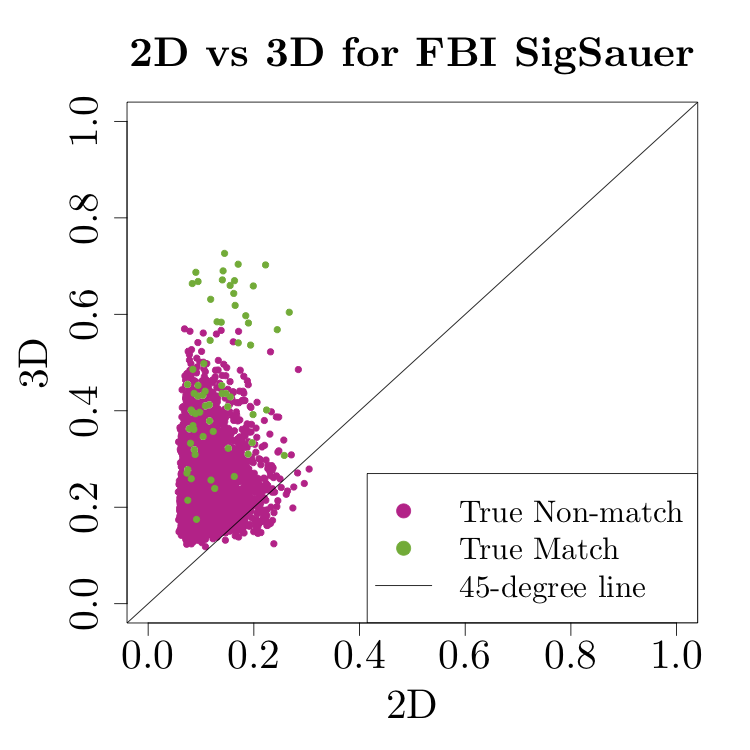}
		\caption{\label{fig:sig2Dvs3D} FBI Sig Sauer}
	\end{subfigure}
	~
	\begin{subfigure}[t]{0.23\textwidth}
		\centering
		\includegraphics[width=\columnwidth]{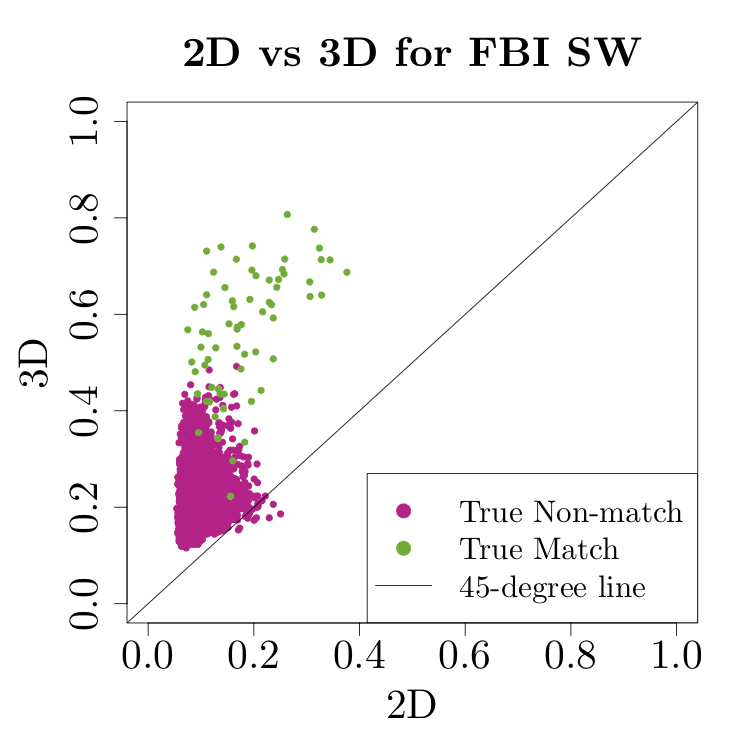}
		\caption{\label{fig:sw2Dvs3D} FBI Smith \& Wesson}
	\end{subfigure}
	~
	\begin{subfigure}[t]{0.23\textwidth}
		\centering
		\includegraphics[width=\columnwidth]{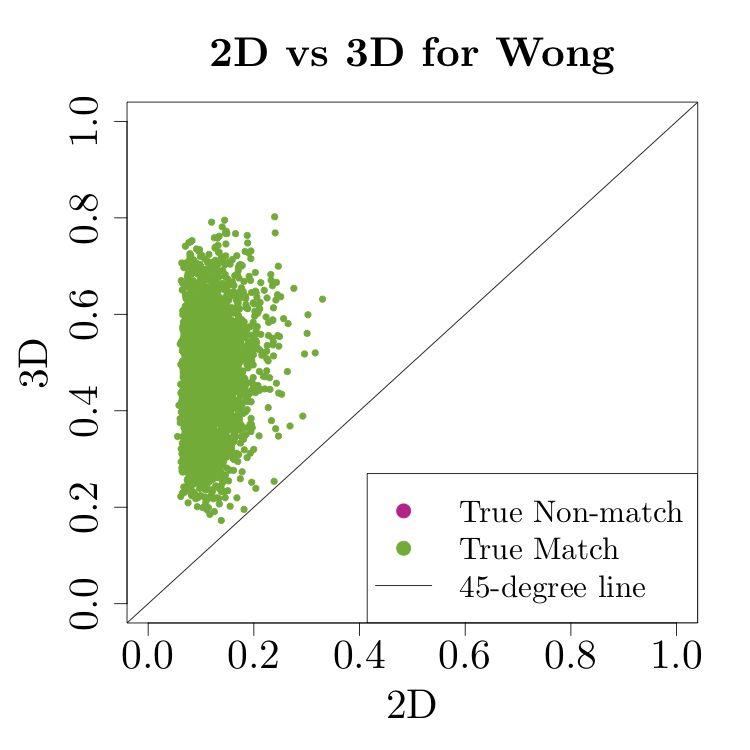}
		\caption{\label{fig:wong2Dvs3D} Cary Wong}
	\end{subfigure}

\end{figure}

The individual correspondences of 2D and 3D scores for each pairwise comparison is in Figure \ref{fig:2Dvs3D}. We notice that there is no straightforward correspondence between the scores, and it is not uniformly the case that results in 3D are superior. For some of the consecutively manufactured data sets for example, the separation in match and non-match score distributions is actually better in 2D than in 3D. It is true in general however, that 3D produces better performance. 3D scores are generally higher than 2D scores, and high 2D scores generally correspond to high 3D scores. If the 2D score is low however, 3D scores span the entire range of values. For non-matches this is undesirable, but for matches the opposite is true, resulting in the overall better performance using 3D images.



\section{Discussion and conclusion}
\label{sec:discussion}

In this paper, we develop methodology to match topographical measurements of cartridge cases. This methodology is both fully automatic and open source, and we evaluate it comprehensively on over 1,100 publicly-available images. We compare results to that using older reflectance microscopy images.

We make several specific improvements over existing methods. First, the selection of breechface marks is automated using RANdom SAmple Consensus (RANSAC). Removing circular symmetry during pre-processing removes artifacts like the slope of the primer surface downwards towards the center. This reduces the likelihood of spurious correlations. We standardize evaluation using precision-recall graphs, and attach a numerical summary by reporting the area under the curve. Finally, resolving intransitive links and generating a linked data set has not been given much thought in the forensics literature, and we propose ways to address this using ideas from the record linkage literature. We contribute a non-traditional application of record linkage.

Several domain-specific conclusions that are of interest are that performance using topographical measurements have better overall performance than reflectance images over a large number of data sets, but individual comparisons vary in quality. Excellent overall performance is achieved (area under the precision-recall curve of over .9) in 9 out of the 13 data sets studied. Performing the clustering step generally improved performance, particularly for data sets with middling to good (but not excellent) results. Performance is generally good on consecutively manufactured data sets, whereas other types of data sets could have much poorer performance. It is not possible to make general statements about particular gun or ammunition brands, since there is a lot of variability depending on the gun model and data collection effort. One clear conclusion from the variety of data sets analyzed, is that some specific firearm and ammunition combinations produce far superior results than others. More work needs to be done to investigate and explain these differences. 


There are at least two specific directions of future work that are essential if such methods are to be adopted for use in actual criminal cases. The first is to do a much more comprehensive comparison of the results of automatic methods with that of human examiners. There is some evidence to suggest that examiner performance might be related to algorithmic performance; for example, \cite{Lightstone2010} concludes that examiners did not have difficulty differentiating consecutively manufactured slides, and in this paper neither does an automatic method. Much more work needs to be done before reaching a definitive conclusion, but establishing a relationship between examiner and algorithmic performance would be beneficial in many ways. For example, data on examiner error rates are generally lacking, and results using automatic methods may serve as a guide to examiner error rates. Types of comparisons that are problematic for algorithms could suggest problems for examiners as well. 
The second issue is to examine the fairness of automatic methods. As such methods have risen in popularity in forensic matching, there have been calls from government to consider this issue \citep{Takano:2019aa}. In firearms matching, we have already seen large differences in performance depending on the data set studied. The first step would be do a proper accounting of these differences. Next and more importantly, is to examine if these translate into bias towards any particular group, for example through geography, or different costs of guns and ammunition resulting in different groups favoring certain types of guns or ammunition. 





To conclude, automatic methods have a lot of potential in terms of corroborating, enhancing, and possibly one day replacing examiner testimony. We have produced good results on many data sets, but these methods need to be tested much more extensively before being used in real cases, where there is the possibility of unintended harms, and where the consequences of falsely implicating a suspect are severe.




\setlength{\bibsep}{0pt plus 0.3ex}
\bibliographystyle{humannat}
\bibliography{cartridgerefs.bib,refs.bib,fairness.bib}

\end{document}